\documentclass[12pt]{article}
\usepackage{graphicx,color}
\graphicspath{images/}
\usepackage{amsmath}
\usepackage[table]{xcolor}
\usepackage[font=small, labelfont=bf]{caption}
\usepackage{subfigure}
\usepackage{subfiles} 

\usepackage{dcolumn}
\usepackage{capt-of}
\usepackage{bm}
\usepackage{tabularx,booktabs}

\newcommand{\be}{\begin{equation}}
\newcommand{\ba}{\begin{eqnarray}}
\newcommand{\ee}{\end{equation}}
\newcommand{\ea}{\end{eqnarray}}

\begin{document}
\title{Rationally Extended Harmonic Oscillator potential, Isospectral Family
and the Uncertainity Relations}
	
\author{Rajesh Kumar$^{a}$\footnote{e-mail address: kr.rajesh.phy@gmail.com(R.K)}, Rajesh Kumar Yadav$^{b}$\footnote{e-mail address: rajeshastrophysics@gmail.com(R.K.Y)} and 
Avinash Khare$^{c}$\footnote {e-mail address: avinashkhare45@gmail.com (A.K)}}
 \maketitle
 
{$~^a$Department of Physics, Model College, Dumka-814101, India.\\
$~^b$Department of Physics, S. K. M. University, Dumka-814110, India.\\
$~^c$Department of Physics, Savitribai Phule Pune University, Pune-411007, India. }

\begin{abstract}
We consider the rationally extended harmonic oscillator potential which is 
isospectral to the conventional one and whose solutions are associated with the
exceptional, \(X_m\)- Hermite polynomials and discuss its various important 
properties for different even codimension of $m$. The uncertainty relations are
obtained for different \( m \) and it is shown that for the ground state, the 
uncertainity increases as \(m\) increases.
 A one parameter $(\lambda)$ family  of exactly solvable isospectral potential 
corresponding to this extended harmonic oscillator potential is obtained. 
Special cases corresponding to the $\lambda=0$ and $\lambda = -1$, which give 
the Pursey and the Abhram-Moses potentials respectively, are discussed. The 
uncertainty relations for the entire isospectral family of potentials for different $m$ and $\lambda$ are also calculated.
\end{abstract}


\section{Introduction}
The idea of Supersymmetric Quantum Mechanics(SQM) 
\cite{cooper1995supersymmetry} is not only useful in solving the quantum 
mechanical potential problems but has also opened the scope for discovering new
exactly solvable potentials. These potentials have applications in diverse 
areas like inverse scattering \cite{agranovich2020inverse, chadan2012inverse}, 
soliton theory \cite{lamb1980elements, drazin1989solitons}, etc. This sparked a
race among researchers to search for a family of isospectral potentials 
\cite{nieto1984relationship, amado1988phase, pursey1986isometric, 
abraham1980changes}. To accomplish this purpose, several methods were developed
like Darboux transformation \cite{darboux1999proposition}, Darboux Crum Krein 
Adler Transformation \cite{odake2013krein}, SQM \cite{khare1989phase, 
keung1989families}, etc. Popular among them was the SQM approach due to its 
simplicity and it was shown using this approach that for any 1-D potential with 
atleast one bound state, one can always construct one continuous parameter 
family of strictly isospectral potentials.

The discovery of the \( X_m \)-exceptional orthogonal polynomials (EOPs) 
\cite{gomez2009extended, gomez2010exceptional, gomez2012orthogonal} paved the 
path for discovering new rationally extended shape invariant 
potentials whose eigenfunctions are in terms of these EOPs. Various properties 
of these potentials have been studied in detail in \cite{midya2009exceptional, ho2011dirac, yadav2013scattering, yadav2015scattering, ho2014scattering, yadav2015group, kumari2016scattering, yadav2016parametric, kumari2017class, 
yadav2019rationally, basu2017quasi, banerjee2022solutions, ramos2017short} and the references 
therein. After the discovery of the exceptional Hermite polynomials 
\cite{gomez2013rational}, Fellows and Smith 
\cite{fellows2009factorization} discovered rationally extended one 
dimensional harmonic oscillator potentials. Their work has been further 
extended using the SQM 
approach \cite{marquette2013two}.

The one dimensional harmonic oscillator potential is one of the most important
potential having numerous applications. However, so far as we are aware off, 
there has not been much progress in studying the various properties of the 
rationally extended family of harmonic oscillator (REHO) potentials. The 
purpose of this paper is to take a step in that direction. Firstly, we 
calculate the Heisenberg uncertainty relation $\Delta x \Delta p$ for the
REHO potentials. Further, we follow the idea of SQM 
\cite{cooper1995supersymmetry}, and generate
a one parameter $(\lambda)$ family of rationally extended strictly isospectral 
potentials including the corresponding Pursey and the Abrahm-Mosses potentials
and obtain their eigenfunctions explicitly in terms of the $X_m$-Hermite EOPs. We calculate the Heisenberg uncertainty relations for the one 
parameter family of rationally extended isospectral potentials (including the
corresponding Pursey and the Abrahm-Mosses potentials) for different $m$ and 
$\lambda$.

The plan of the paper is as follows: In Sec. $2$, we briefly discuss the 
formulation of SQM. In Sec. $3$, we summarise the known important results 
related to the rationally extended harmonic oscillator potentials. 
A one parameter $\lambda$ family of isospectral potentials (including the 
corresponding Pursey and Abrahm-Mosses potentials) are obtained in 
Sec. $4$ for any even integral $m$. In Sec. $5$, we follow the results 
discussed in Sec. $3$ and Sec. $4$ and calculate the Heisenberg uncertainty 
relations for REHO and their isospectral family of potentials (including the
corresponding Pursey and Abrahm-Mosses potentials). Finally, we 
summarize our results and mention some open possible problems in Sec. $6$.


\section{SQM Formalism}
In SQM approach, one considers the Hamiltonian (in the units $\hbar=2m=1$)
\begin{equation}
H^-= -\frac{d^2}{dx^2}+V^-(x)-\epsilon\; \label{H-}
\end{equation}
where $\epsilon$ is the factorization energy. By assuming the ground state 
energy of this Hamiltonian $E^-_0=0$, we factorize $H^-$ in terms of  $ A$ and 
$A^{\dagger}$ as 
\begin{equation}
H^-= A^{\dagger}A\;
\end{equation}
with
\begin{equation}
A=\frac{d}{dx}+W(x),\qquad A^{\dagger}=-\frac{d}{dx}+W(x). \label{A-form}
\end{equation} 
Here $W(x)$ is known as superpotential which is expressed in term of the ground
state eigenfunction  as $W(x)=-\ln[\psi^-_0(x)]'$. In this way, another set of 
Hamiltonian can easily be constructed by reversing the order of the operators 
$A$ and $A^\dagger$ i.e., 
\begin{equation}
H^+=AA^{\dagger}=- \frac{d^2}{dx^2}+V^+(x)-\epsilon\;. \label{H+}
\end{equation}
Thus, the partner potentials \( V^{\mp}(x) \) in term of superpotential are given by 
\begin{equation}
V^{\mp}(x)=W(x)^2\mp W(x)'+\epsilon  \;. \label{V-form}
\end{equation}
Here a prime denotes a derivative with respect to \( x \).
The precise relationship between the energies and the eigenfunctions of the 
partner Hamiltonians are 
\begin{equation}
E^-_{n+1}=E^+_n\;> 0,\qquad n=0,1,2,\cdots,
\end{equation}
\begin{equation}
\psi^-_{n+1}(x)= \sqrt{\frac{1}{E^+_n}} A^{\dagger} \psi^+_n(x),\qquad\label{es}
\psi^+_n(x)= \sqrt{\frac{1}{E^-_{n+1}}} A \psi^-_{n+1}(x).
\end{equation}
The ground state eigenfunction \( \psi^-_0(x) \) is obtained by solving the differential equation,
\begin{equation}
A \psi^-_{0}(x)=0.
\end{equation}

One parameter family of isospectral potentials are obtained by redefining the 
form of the superpotential 
\begin{equation}
\hat{W}=W(x)+\phi(x),
\end{equation}
and by assuming the uniqueness of the partner potential $V^+(x)$ i.e.,
\begin{equation*}
V^+(x)=\hat{W}(x)^2+\hat{W}(x)' +\epsilon=W(x)^2+W(x)'+\epsilon,
\end{equation*}
which gives 
\begin{equation*}
\phi^2(x)+2W\phi+\phi'(x)=0.
\end{equation*}
We then find that on substituting \( \phi(x)= y^{-1}(x)\), the above equation 
satisfies the Bernoulli equation
\begin{equation*}
y'(x)=1+2W(x)y(x),
\end{equation*}
whose solution is 
\begin{equation*}
\phi(x)= \frac{d}{dx}\ln\left[\mathcal{I}(x)+\lambda\right].
\end{equation*}\\
Here \( \mathcal{I}(x)=\int^{x}_{\infty} \psi^-_{0}(x')^2\;dx' \) and $\lambda$ is an integration constant.\\

Therefore, the potential which is strictly isospectral to \( V^-(x) \) is 
given by 
\begin{align}
\hat{V}^-(\lambda, x)&= \hat{W}(x)^2- \hat{W}(x)' +\epsilon \nonumber\\
&=V^-(x)-2 \frac{d^2}{dx^2}\ln\left[\mathcal{I}(x)+\lambda\right]\,,
\end{align}
where either $\lambda > 0$ or $\lambda < -1$ so as to avoid singularity
(For details, see\cite{cooper1995supersymmetry, yadav2022one}). The 
normalized ground state eigenfunction for the entire family of potentials
is given by
\begin{equation}
\hat{\psi}^-_{0}(\lambda, x)= \sqrt{\lambda(1+\lambda)}\;\frac{\psi^-_{0}(x)}
{\mathcal{I}(x)+\lambda}\,.\label{iso-gs}
\end{equation}\\
The normalized excited states eigenfunctions can be easily calculated similar to (\ref{es}) as
\begin{align}
\hat{\psi}^-_{n+1}(\lambda,\;x)&= \sqrt{\frac{1}{E^+_n}} \hat{A}^{\dagger}\psi^+_n(x)=\frac{1}{E^-_{n+1}} \hat{A}^{\dagger}A \psi^-_{n+1}(x)\label{iso-es},\qquad n=0,1,2,\cdots,\\
\hat{A}^{\dagger}&=-\frac{d}{dx}-\frac{d}{dx}\ln\left[\frac{\psi^-_{0}(x)}{\mathcal{I}(x)+\lambda}\right], \qquad A =\frac{d}{dx}-\frac{d}{dx}\ln\left[\psi^-_{0}(x)\right]\nonumber
\end{align}
and the Hamiltonian is defined similar to (\ref{H-}) as 
\begin{align}
\hat{H}^-&=\hat{A}^{\dagger} \hat{A}= - \frac{d^2}{dx^2}+\hat{V}^-(x)-\epsilon\label{Hiso-}\\
\hat{A}&=\frac{d}{dx}-\frac{d}{dx}\ln\left[\frac{\psi^-_{0}(x)}{\mathcal{I}(x)+\lambda}\right]\nonumber
\end{align}

It is worth reminding that all these strictly isospectral family of potentials 
have same partner potential $V^+(x)$.

In the limit \( \lambda=0 \) there is a loss of a bound state and the 
corresponding potential is called the Pursey potential \( \hat{V}^{P}(x) \). 
An analogous situation occurs in the limit \( \lambda=-1 \) and the potential 
is called the Abraham-Moses potential \( \hat{V}^{AM}(x) \). 
The normalized eigenfunctions of \( \hat{V}^{P}(x) \) are given by
\begin{equation}
\hat{\psi}^{P}_{n}(x)=\frac{1}{E^-_{n+1}} \left(-\frac{d}{dx}-\frac{d}{dx}\ln\left[\frac{\psi^-_{0}(x)}{\mathcal{I}(x)}\right]\right)A\psi^-_{n+1}(x),\qquad n=0,1,2,\cdots,\label{P}.
\end{equation}

Similarly, the normalized eigenfunctions of \( \hat{V}^{AM}(x) \) are given by
\begin{equation}
\hat{\psi}^{AM}_{n}(x)=\frac{1}{E^-_{n+1}} \left(-\frac{d}{dx}-\frac{d}{dx}\ln\left[\frac{\psi^-_{0}(x)}{\mathcal{I}(x)-1}\right]\right)A\psi^-_{n+1}(x),\qquad n=0,1,2,\cdots,\label{AM}.
\end{equation}
The energy eigenvalues of the Pursey and the Abraham-Moses Hamiltonians obtained by substituting \( \lambda \) equal to \( 0 \) and \( -1 \) respectively in (\ref{Hiso-}) has the same expressions as that of $H^{+}(x)$, i.e.
\begin{equation}
\hat{E}^{P}_{n}=\hat{E}^{AM}_{n}=E^+_{n},\qquad n=0,1,2,3,\cdots\label{E-pam}
\end{equation}

\section{REHO Potential}
In this section, we briefly review the results obtained in  
\cite{ fellows2009factorization,marquette2013two} about the REHO 
potentials. These authors extended the conventional one-dimensional harmonic 
oscillator potential ($V(x)=x^2$) using the idea of SQM and obtained the REHO 
potentials valid for even co-dimension of $m=0,2,4,....$ and are given by
\begin{equation}
V^-_m(x)=V(x)-2\left[ \frac{\mathcal{H}_m''}{\mathcal{H}_m}-\left(\frac{\mathcal{H}_m'}{\mathcal{H}_m}\right)^2+1\right],
\end{equation} 
where $\mathcal{H}_m(x)=(-1)^mH_m(ix)$ is pseudo-hermite polynomials and factorization energy used in the calculation was \( \epsilon=-2m-1 \). The 
normalized ground state eigenfunction \( \psi^-_{0,m}(x) \) having zero 
eigenvalue and the excited states eigenfunction\( \psi^-_{n+1,m}(x) \) for different $m$ are
\begin{equation}
\psi^-_{0,m}(x)=\left( \frac{2^m m!}{\sqrt{\pi}}\right)^{ \frac{1}{2}}  \frac{e^{- \frac{x^2}{2}}}{\mathcal{H}_m(x)}, \label{RE-GS}
\end{equation} 
and
\begin{equation}
\psi^-_{n+1,m}(x)=\frac{1}{\sqrt{E^+_{n,m}\;2^{n}n!\sqrt{\pi}}}\;\frac{e^{- \frac{x^2}{2}}}{\mathcal{H}_m(x)} y^m_{n+1}(x) \;,\qquad n=0,1,2...,	
\end{equation}
 respectively. 
Here \( y^m_{n+1}(x)=\left[\mathcal{H}_m(x)H_{n+1}(x)
+\mathcal{H}_m(x)'H_{n}(x)\right]\) and \( n = -1,\;0,\;1,\;2,\dots \)  
is the Exceptional Hermite Polynomial. Note that 
\(y^{m}_{0}(x) = 1 \).
The whole system $\{ y^m_{n+1}(x)\}$ is the exceptional orthogonal polynomial system, \( X_m \), of co-dimension \( m \) and is orthogonal and complete with respect to the positive-definite measure \( \frac{e^{-x^2}}{\mathcal{H}_m(x)^2} \). 
The Hamiltonian \( H^-_m \) defined similar to (\ref{H-}) has the energy eigenvalues \( E^-_{n,m} \) given by
 \begin{equation}\label{1}
 E^-_{n+1,m}=2(n+m+1),\quad n=0,1,2,\cdots,\quad \text{and } E^-_{0,m}=0. 
 \end{equation} 
It is seen that the difference in energy eigenvalues is \( 2(m+1) \) units between ground state and first excited state and 2 units between any successive excited states. Therefore energy spectra is equidistant only for \( m=0 \). Note that \( V^-_m(x)\) is singular at \( x=0 \) for odd \( m \) and therefore \( m \) is restricted to positive even integers only.

\begin{figure}
\centering
\subfigure[\bf m=0 ]{\includegraphics[height=8cm, width=13cm]{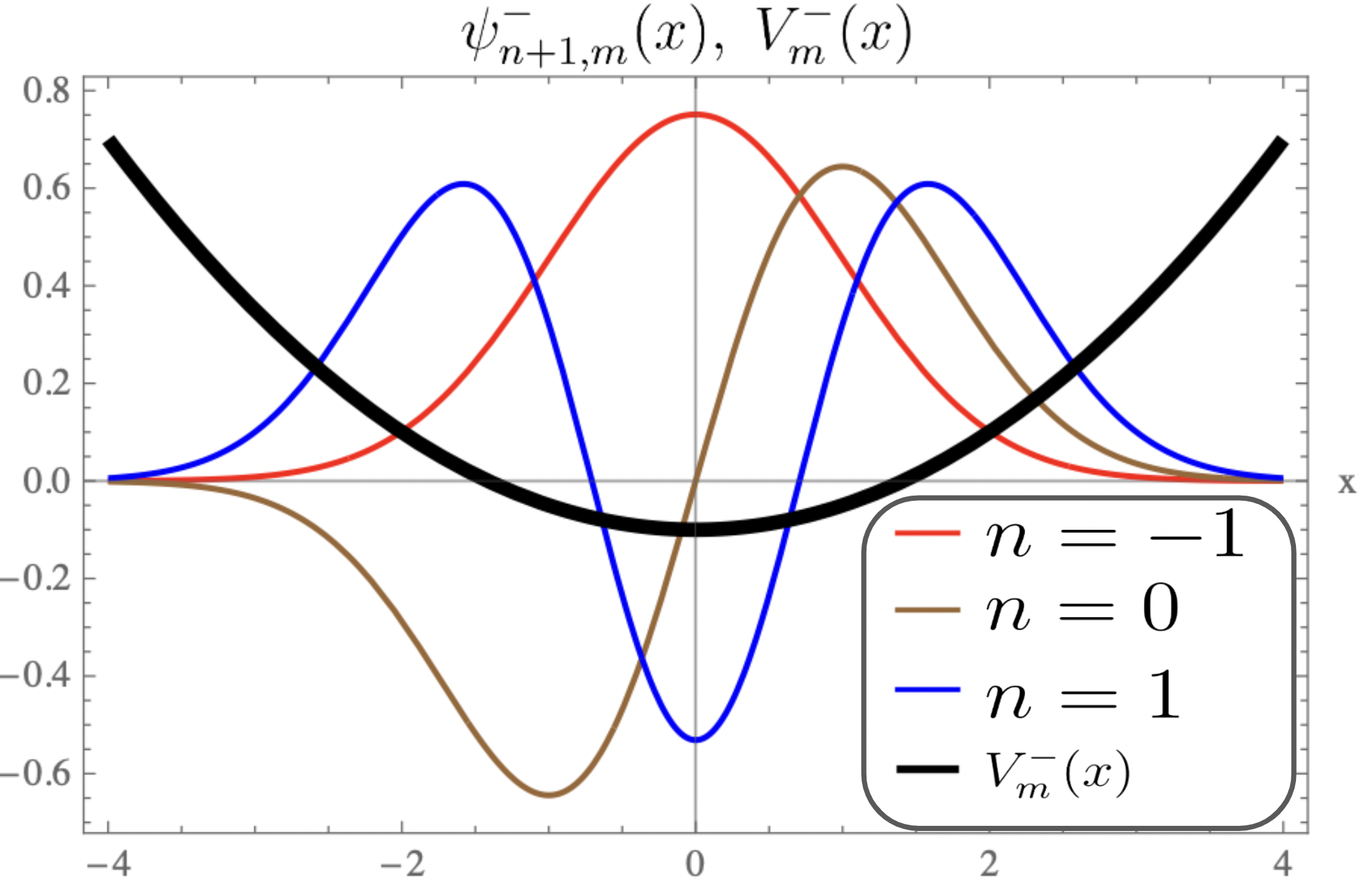}}
\subfigure[\bf m=2 ]{\includegraphics[height=6cm, width=7.9cm]{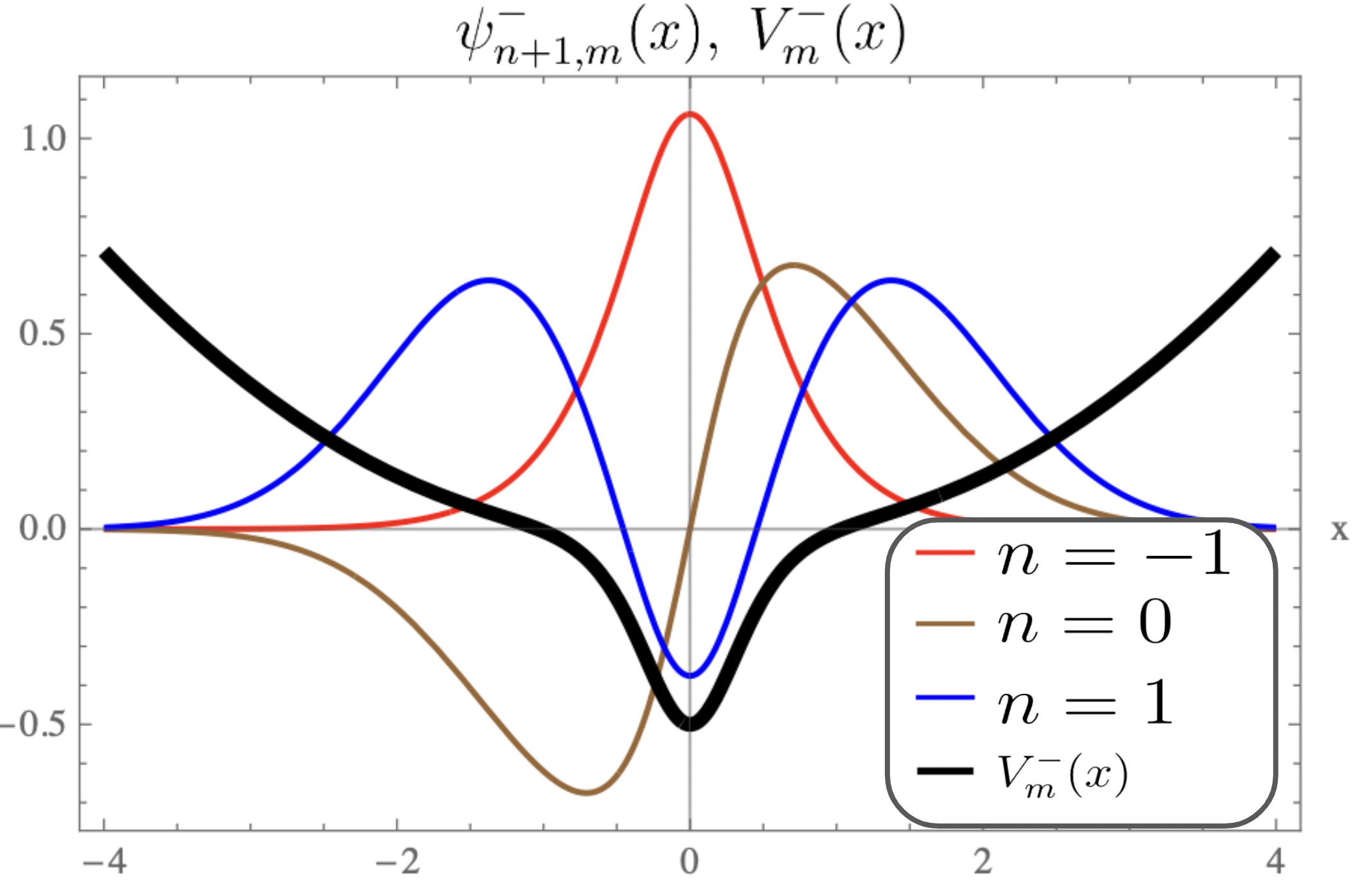}}
\hfill
\subfigure[\bf m=4 ]{\includegraphics[height=6cm, width=7.9cm]{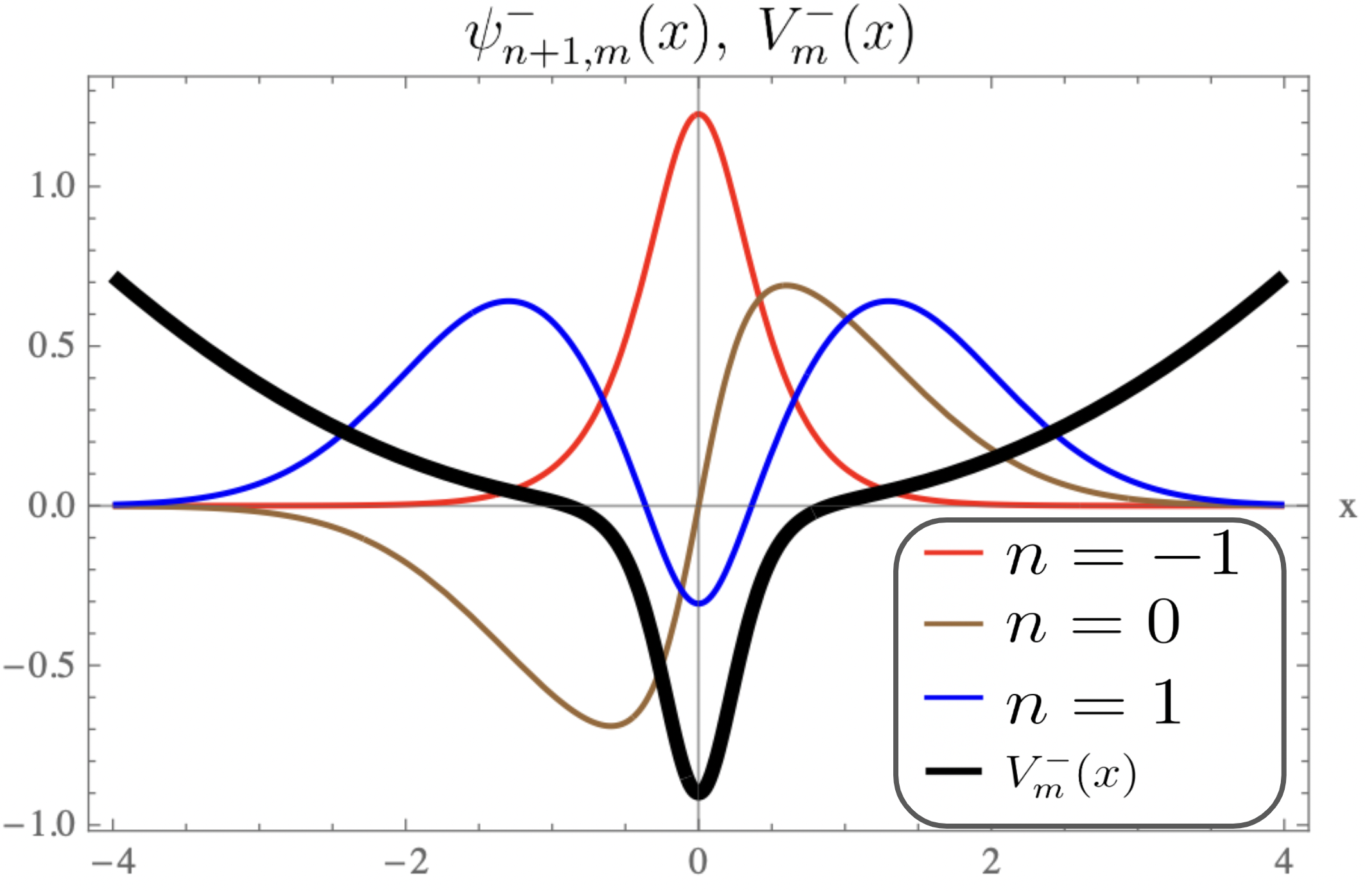}}
\caption{Plots of potentials \(V^-_m(x) \) and eigenfunction \( \psi^-_{n+1,m}(x)\) as a function of \( x \) for \( m=0,\;2\text{ and }4 \). Potentials are shown in black color.}\label{figure-susy}
\end{figure}

It is worth noting that the complete set of eigenfunctions for the $m = 0$ 
case, i.e. the one dimensional harmonic oscillator, can be reduced to a single formula containing ground state eigenfunction and excited state eigenfunction given by
\begin{equation*}
\Psi_{\nu}(x)=\frac{1}{\sqrt{2^{\nu}\nu!\sqrt{\pi}}} e^{-\frac{x^2}{2}} 
H_{\nu}(x) ,\qquad\nu=0,\;1,\;2,\cdots\,.
\end{equation*}
The plot of the ground, first and the second excited state eigenfunctions as 
well as the potential versus position for \( m=0,\;m=2 \text{ and }m=4 \) are 
given in Fig-\ref{figure-susy}.a ,Fig-\ref{figure-susy}.b and 
Fig-\ref{figure-susy}.c respectively. It is interesting to note from these 
figures that as $m$ increases, the eigenfunctions and the potential well 
becomes sharper. In Table $1$ we have given expressions for the potential
$V^{-}_{m}(x)$ as well as the ground and the excited state eigenfunctions 
in case $m = 0, 2,4$. Expressions for exceptional Hermite polynomials 
$y^{m}_{n+1}$ are given in Table $2$ in case $n = 0,1,2$ and $m = 0,2,4$.

The superpotential \( W(x) \) corresponding to the REHO potential is easily 
obtained from the ground state eigenfunctions \( \psi^-_{0,m}(x) \) of \( V^-_m(x) \) as
\begin{eqnarray}
W(x)&=&-\ln [\psi^-_{0,m}(x)]'\nonumber\\
&=& x+\frac{\mathcal{H}_m(x)'}{\mathcal{H}_m(x)}.
\end{eqnarray}
The energy eigenvalues of the partner Hamiltonians $H^{\mp}(x)$, defined using (\ref{H-}) and (\ref{H+}), are given by
\begin{align}
E^-_{n+1,m}&=2(n+m+1),\quad n=0,1,2,\cdots,\quad \text{with}\quad E^-_{0,m}=0.\\
E^+_{n,m}&=2(n+m+1),\quad n=0,1,2,\cdots
\end{align}

\begin{table}[htp]
\centering
\setlength{\tabcolsep}{1em}
\begin{tabularx}{\columnwidth}{@{}>{\bfseries}l X X l @{}}
\toprule
m & \(V^-_m(x) \) & \( \psi^-_{0,m}(x) \)& \( \psi^-_{n+1,m}(x),\quad n=0,1,2\cdots\quad\)\\
\toprule
0& \(x^2-2\)&\( \frac{1}{\sqrt[4]{\pi }}e^{-\frac{x^2}{2}} \)& \( \frac{1}{\sqrt{2(n+1)2^nn!\sqrt{\pi}}} e^{-\frac{x^2}{2}}y^0_{n+1}(x)  \)\\
2& \( x^2-2+\frac{8 \left(2 x^2-1\right)}{\left(2 x^2+1\right)^2}\)&\( \frac{2 \sqrt{2}}{\sqrt[4]{\pi }}\frac{e^{-\frac{x^2}{2}}}{4 x^2+2}\) & \(\frac{1}{\sqrt{2(n+3)2^nn!\sqrt{\pi}}} \frac{e^{-\frac{x^2}{2}}}{4 x^2+2}y^2_{n+1}(x)  \)\\
4& \( x^2 -2 +\frac{16 \left(8 x^6+12 x^4+18 x^2-9\right)}{\left(4 \left(x^2+3\right) x^2+3\right)^2}\)&\(\frac{8 \sqrt{6}}{\sqrt[4]{\pi }}\frac{e^{-\frac{x^2}{2}}}{16 x^4+48 x^2+12}\)& \( \frac{1}{\sqrt{2(n+5)2^nn!\sqrt{\pi}}} \frac{e^{-\frac{x^2}{2}}}{16 x^4+48 x^2+12}y^4_{n+1}(x) \)\\
\bottomrule
\end{tabularx}
\caption{Rationally extended harmonic oscillator potential \(V^-_m(x)\), its ground and excited state wavefunctions for \( m=0,2 \) and $4$.}
\label{tab-re}
\end{table}

\newpage

\begin{figure}
\subfigure[\bf Plot of \(\hat{V}^-_0(\lambda, x)\) as a function of \( x \) for positive \(\lambda\).]{\includegraphics[height=6cm, width=8cm]{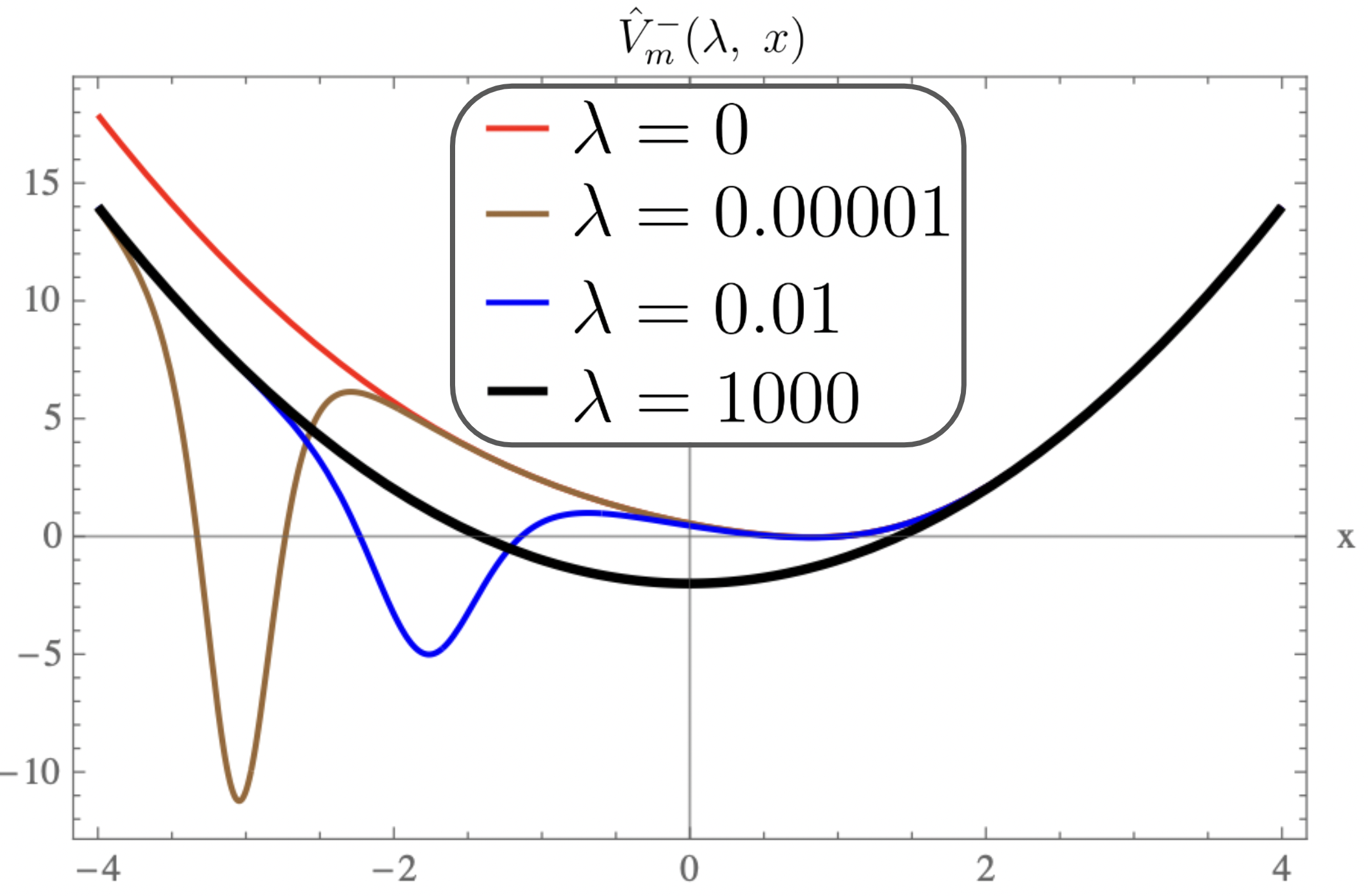}}
\hfill
\subfigure[\bf Plot of \(\hat{V}^-_0(\lambda, x)\) as a function of \( x \) for negative \(\lambda\).]{\includegraphics[height=6cm, width=8cm]{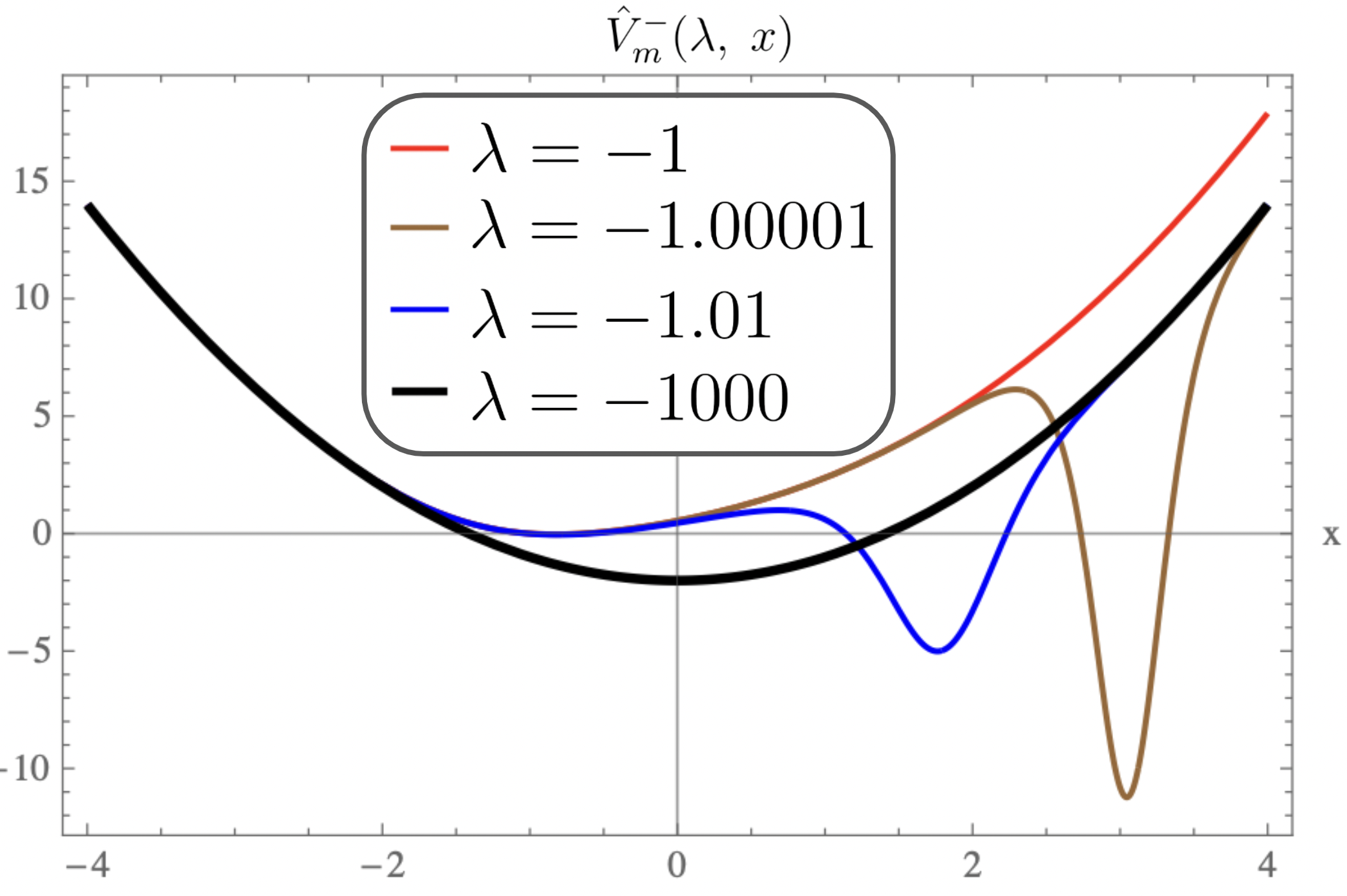}}

\subfigure[\bf Plot of Pursey \(\hat{V}^-_0(0, x)\), AM \(\hat{V}^-_0(-1, x)\) and \(V^+(x)\) potential as a function of \( x \)]
{\includegraphics[height=6cm, width=8cm]{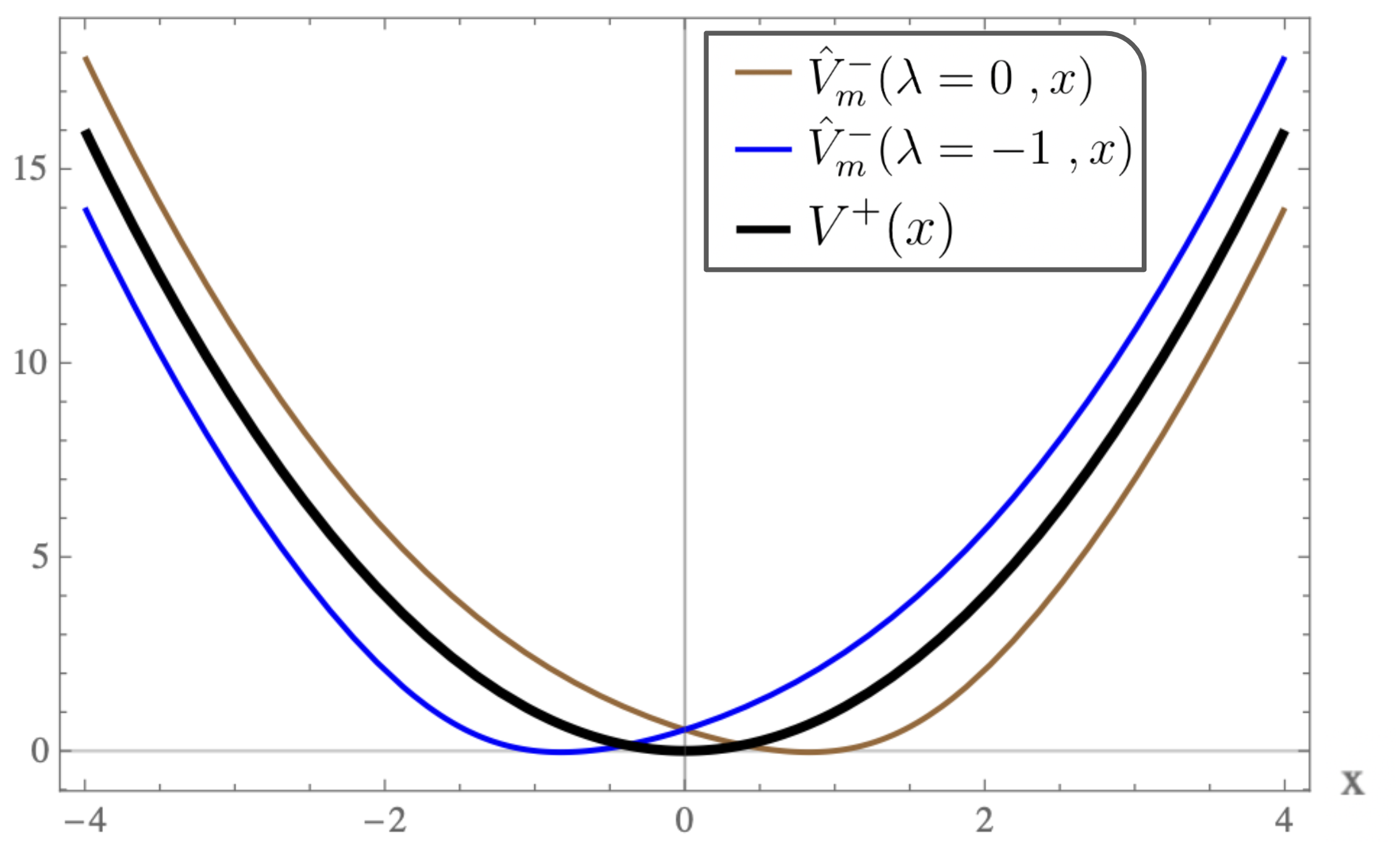}}
\hfill
\subfigure[\bf Plot of Ground-state wavefunction \( \hat{\psi}^-_{0,0}(\lambda,x)\) as a function of \( x \) for various positive \(\lambda\).]{\includegraphics[height=6cm, width=8cm]{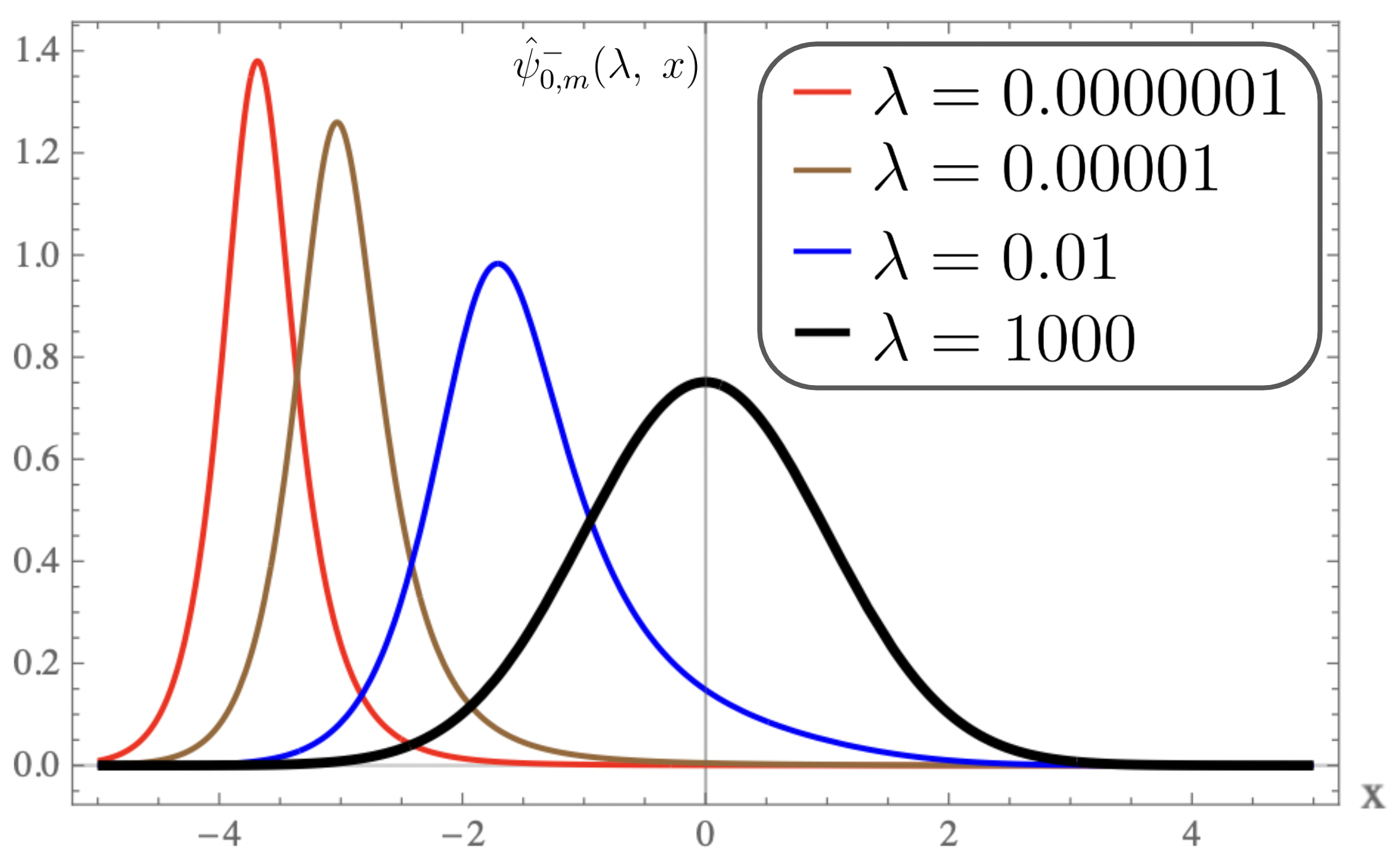}}
\caption{Plot of isospectral potential \(\hat{V}^-_0(\lambda, x)\)  and its ground state eigenfunction \( \hat{\psi}^-_{0,0}(\lambda,x)\) as a function of \( x \) for various \( \lambda \) when \(m=0\).}\label{figure-iso-0}
\end{figure}
\begin{figure}

\subfigure[\bf Plot of \(\hat{V}^-_2(\lambda, x)\) as a function of \( x \) for positive \(\lambda\).]{\includegraphics[height=6cm, width=8cm]{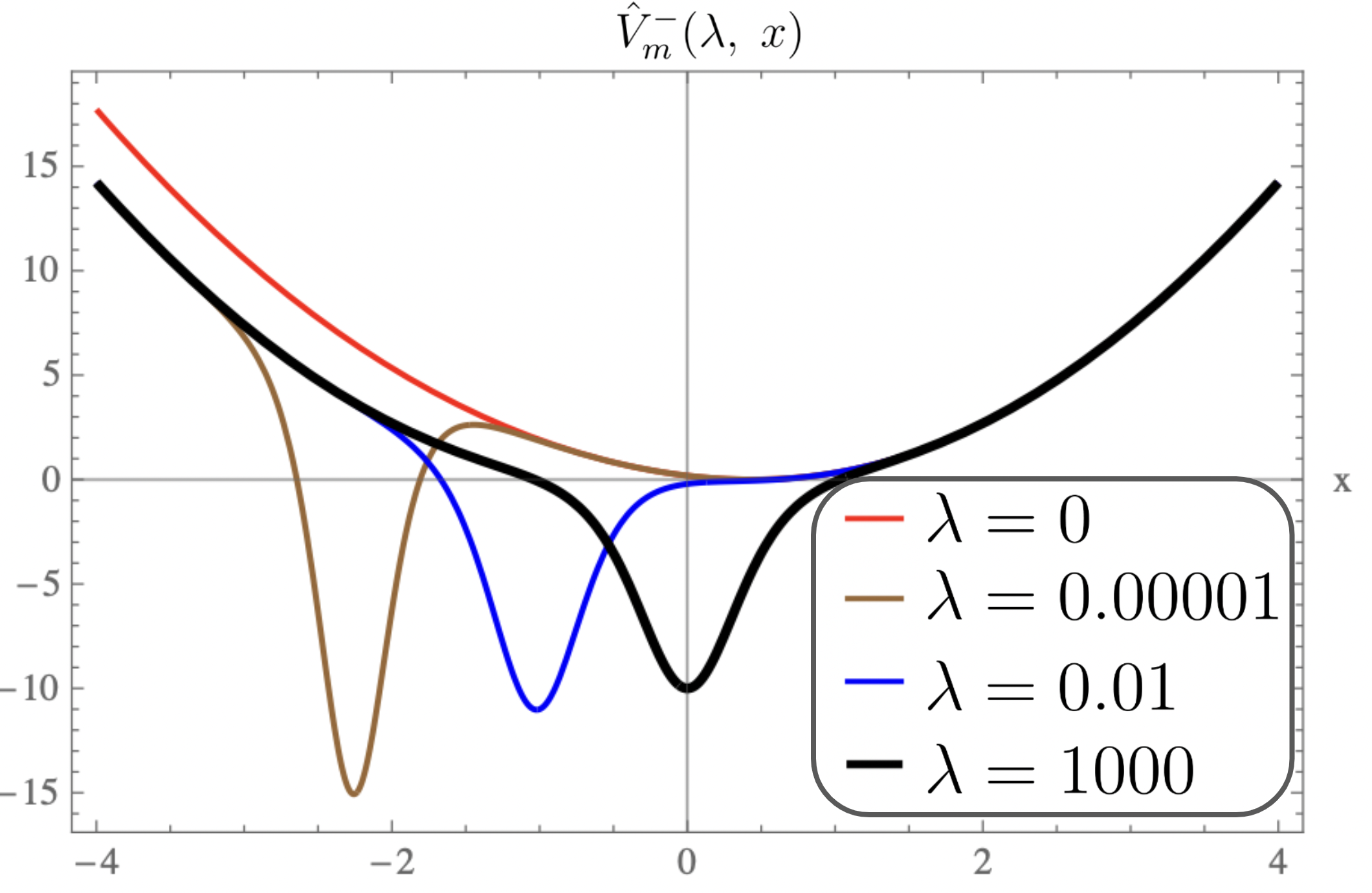}}
\hfill
\subfigure[\bf Plot of \(\hat{V}^-_2(\lambda, x)\) as a function of \( x \) for negative \(\lambda\).]{\includegraphics[height=6cm, width=8cm]{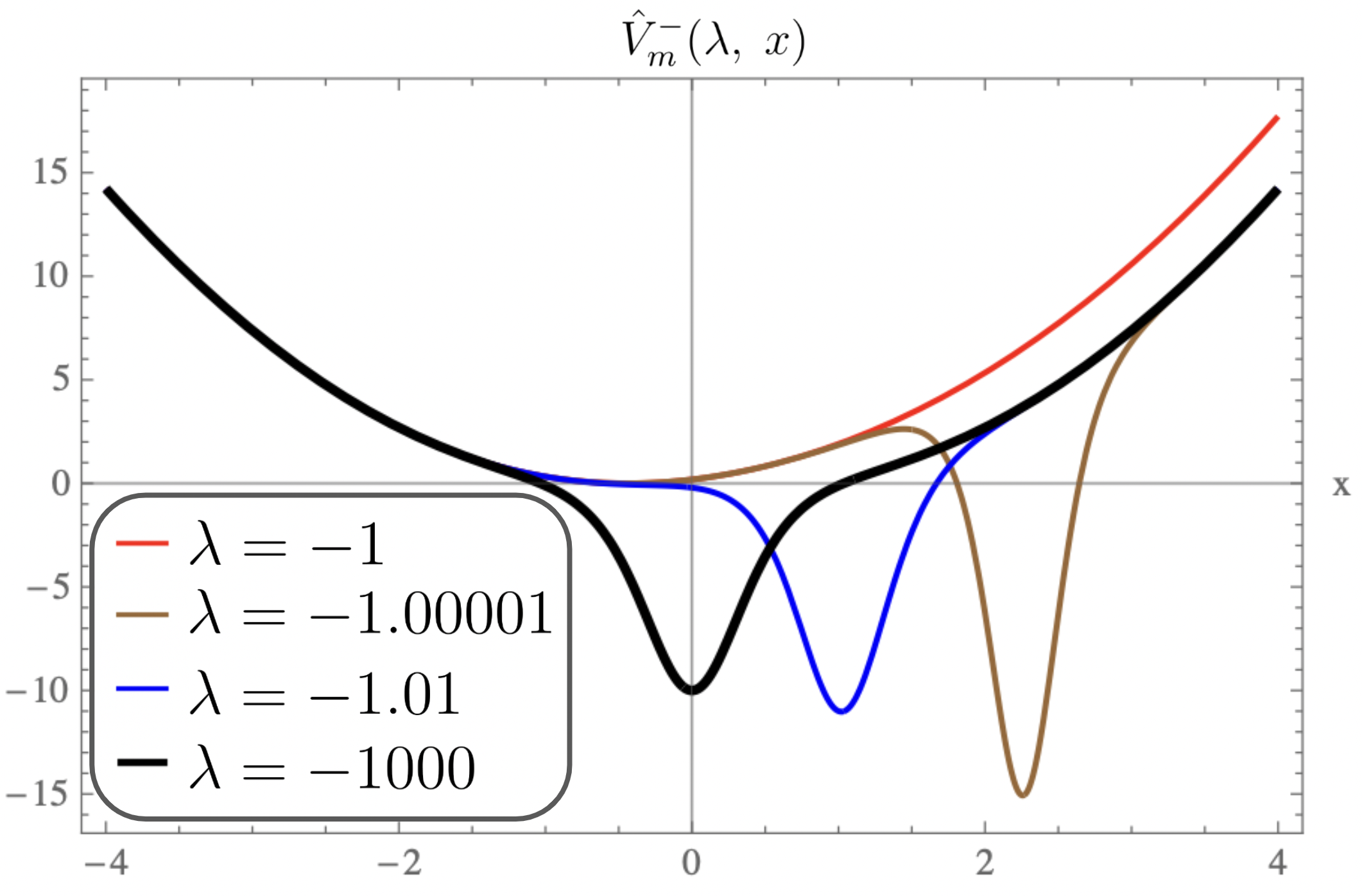}}

\subfigure[\bf Plot of Pursey \(\hat{V}^-_2(0, x)\), AM \(\hat{V}^-_2(-1, x)\) and \(V^+(x)\) potential as a function of \( x \)]
{\includegraphics[height=6cm, width=8cm]{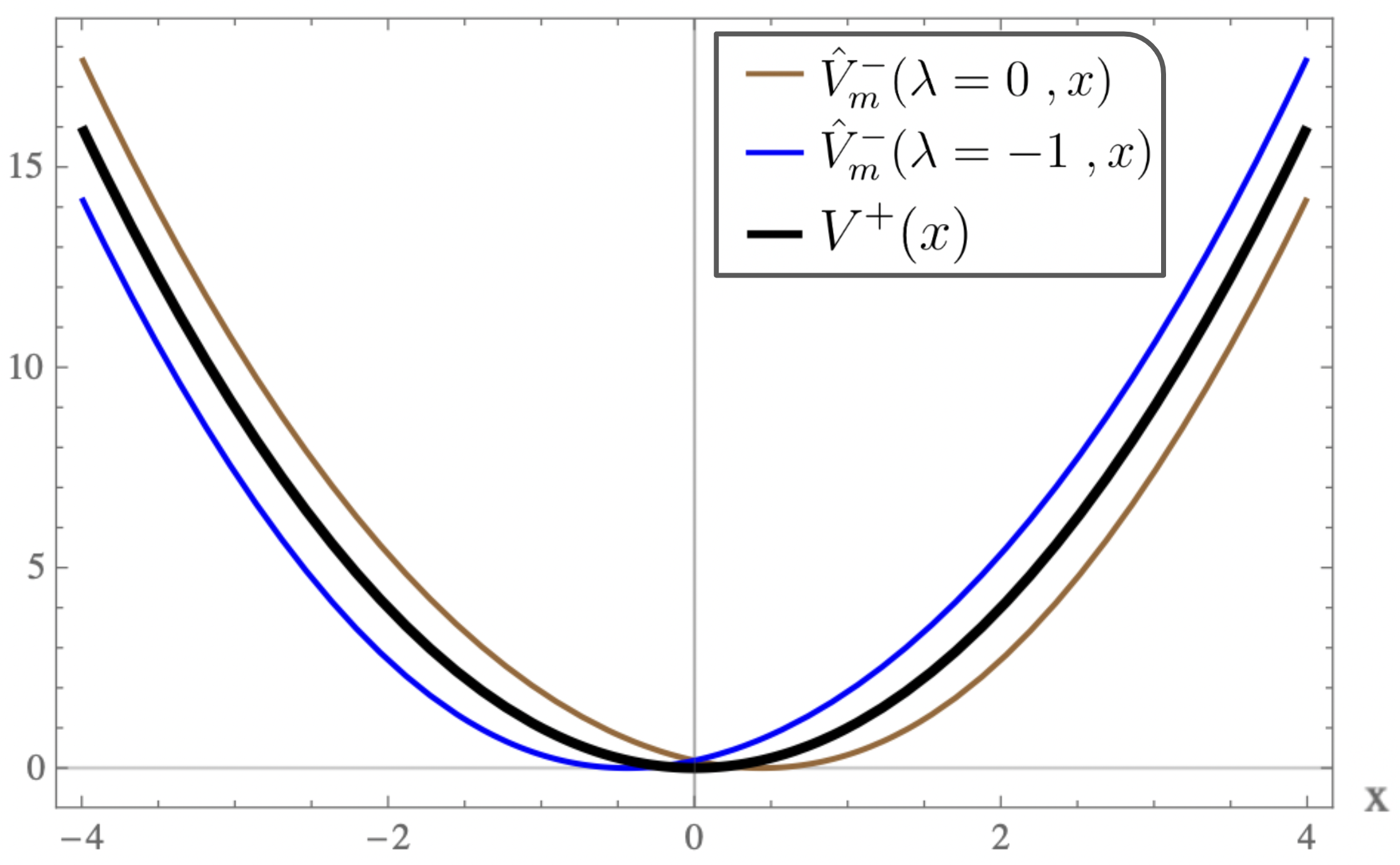}}
\hfill
\subfigure[\bf Plot of Ground-state wavefunction \( \hat{\psi}^-_{0,2}(\lambda,x)\) as a function of \( x \) for various positive \(\lambda\).]{\includegraphics[height=6cm, width=8cm]{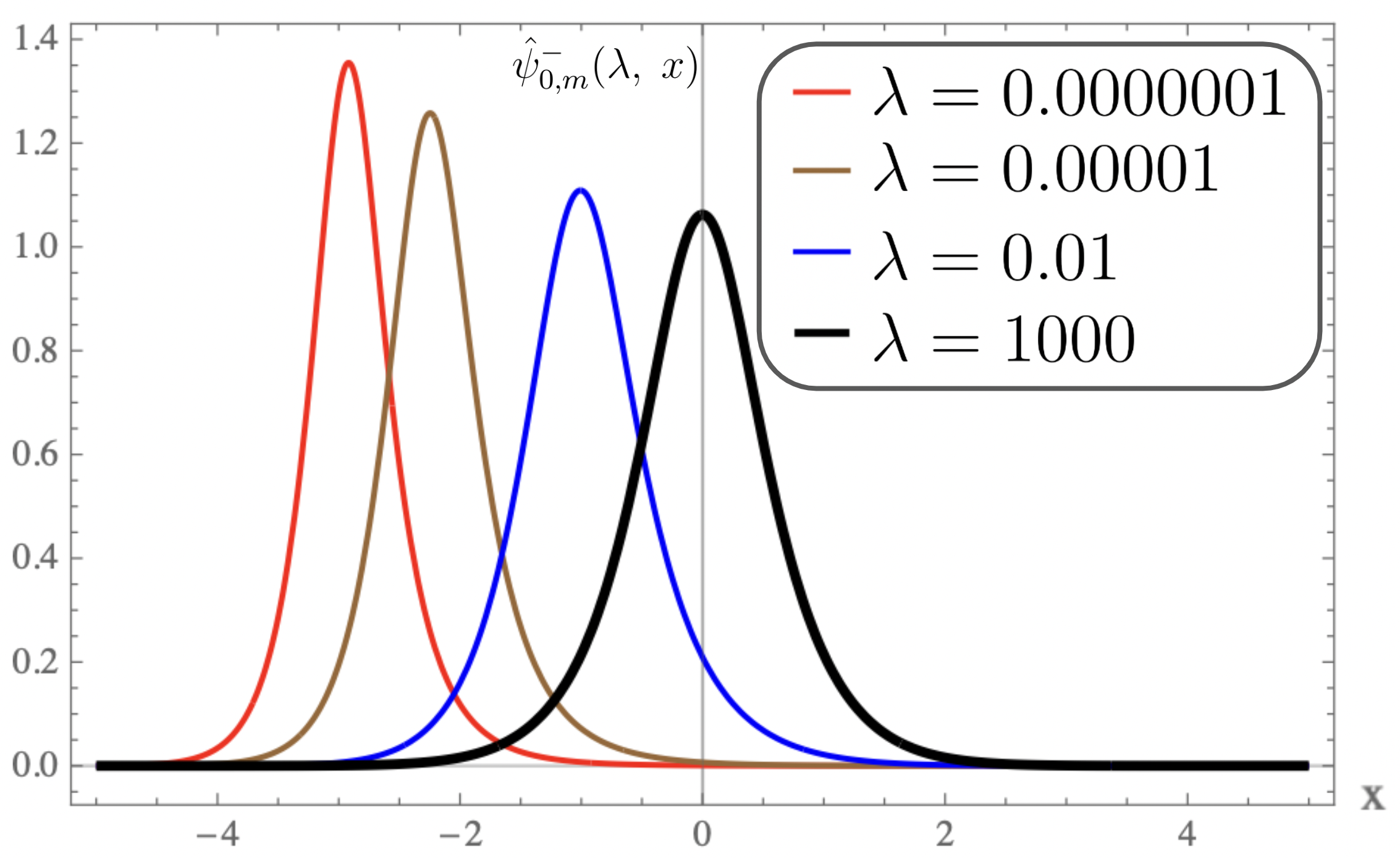}}
\caption{Plot of isospectral potential \(\hat{V}^-_2(\lambda, x)\)  and its ground state eigenfunction \( \hat{\psi}^-_{0,2}(\lambda,x)\) as a function of \( x \) for various \( \lambda \) when \(m=2\).}\label{figure-iso-2}
\end{figure}
\begin{figure}
\subfigure[\bf Plot of \(\hat{V}^-_4(\lambda, x)\) as a function of \( x \) for positive \(\lambda\).]{\includegraphics[height=6cm, width=8cm]{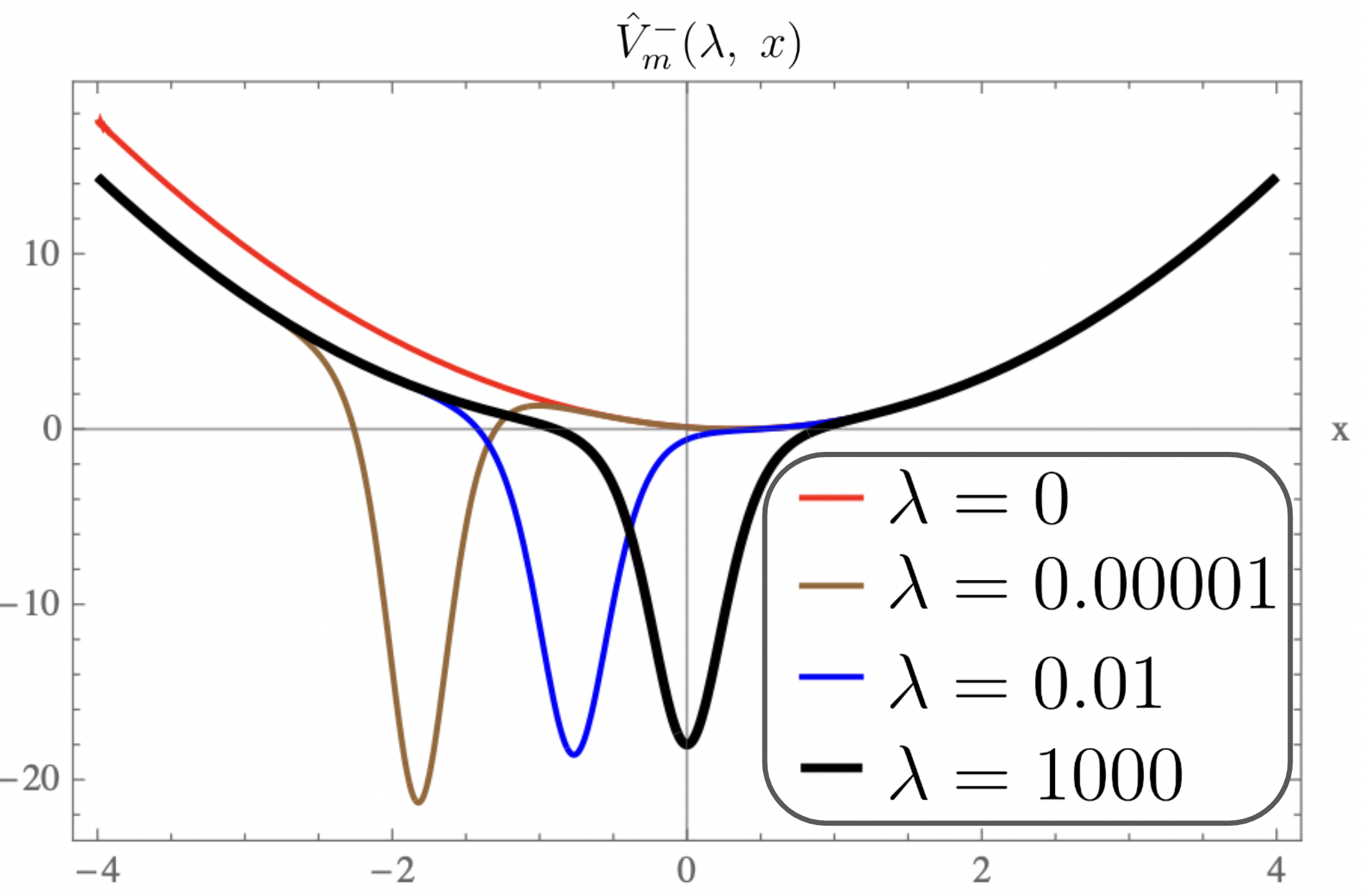}}
\hfill
\subfigure[\bf Plot of \(\hat{V}^-_4(\lambda, x)\) as a function of \( x \) for negative \(\lambda\).]{\includegraphics[height=6cm, width=8cm]{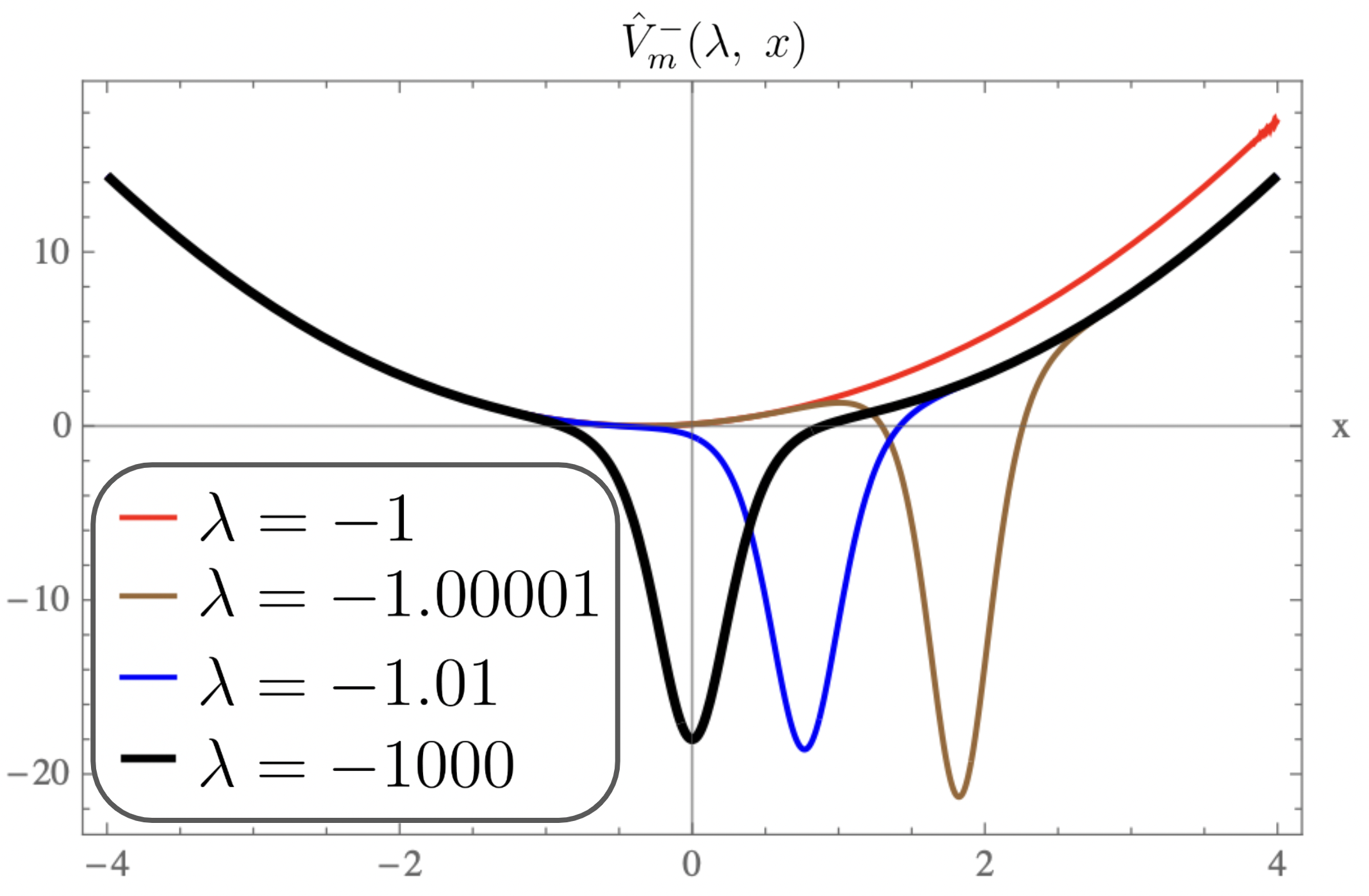}}

\subfigure[\bf Plot of Pursey \(\hat{V}^-_4(0, x)\), AM \(\hat{V}^-_4(-1, x)\) and \(V^+(x)\) potential as a function of \( x \)]
{\includegraphics[height=6cm, width=8cm]{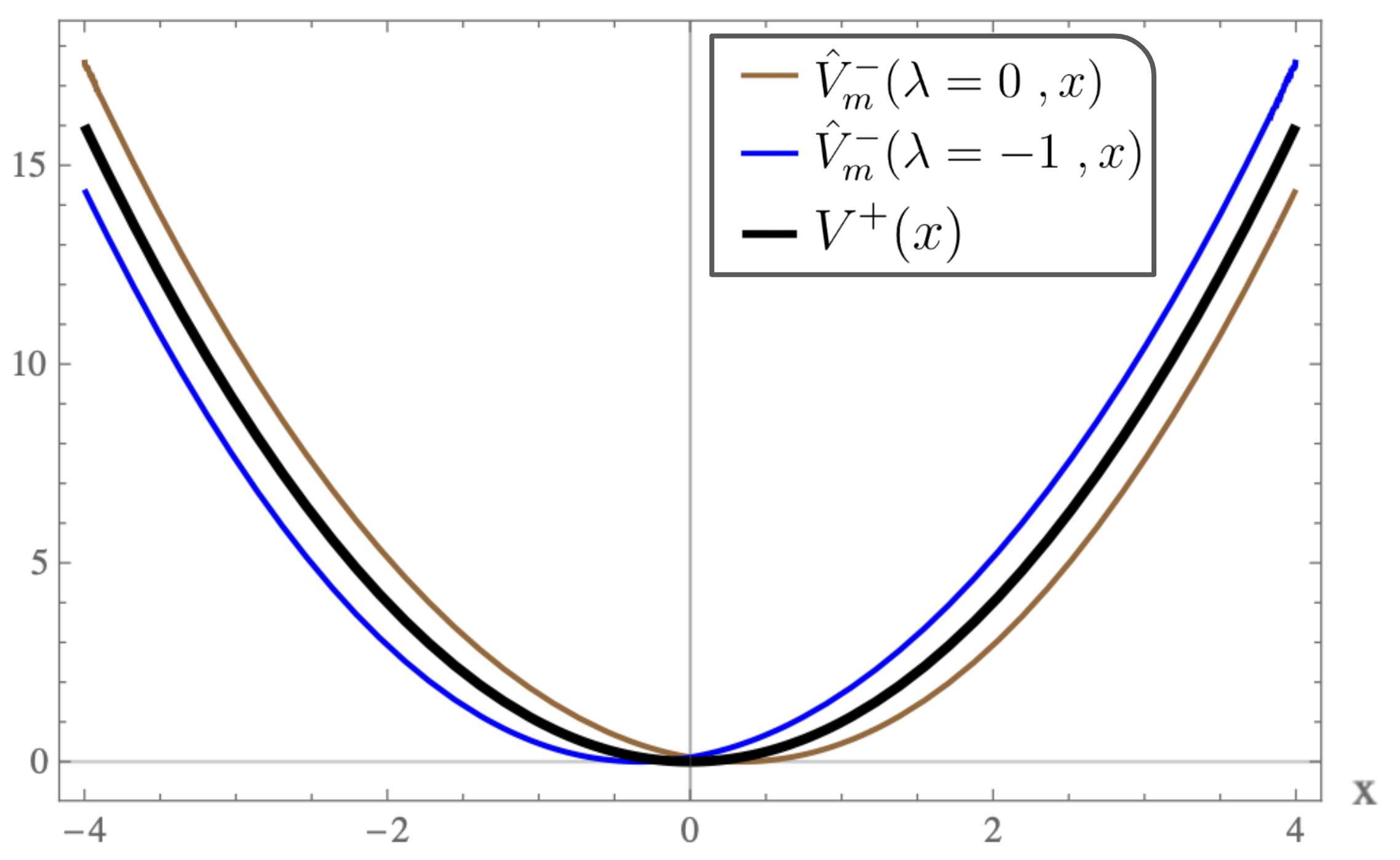}}
\hfill
\subfigure[\bf Plot of Ground-state wavefunction \( \hat{\psi}^-_{0,4}(\lambda,x)\) as a function of \( x \) for various positive \(\lambda\).]{\includegraphics[height=6cm, width=8cm]{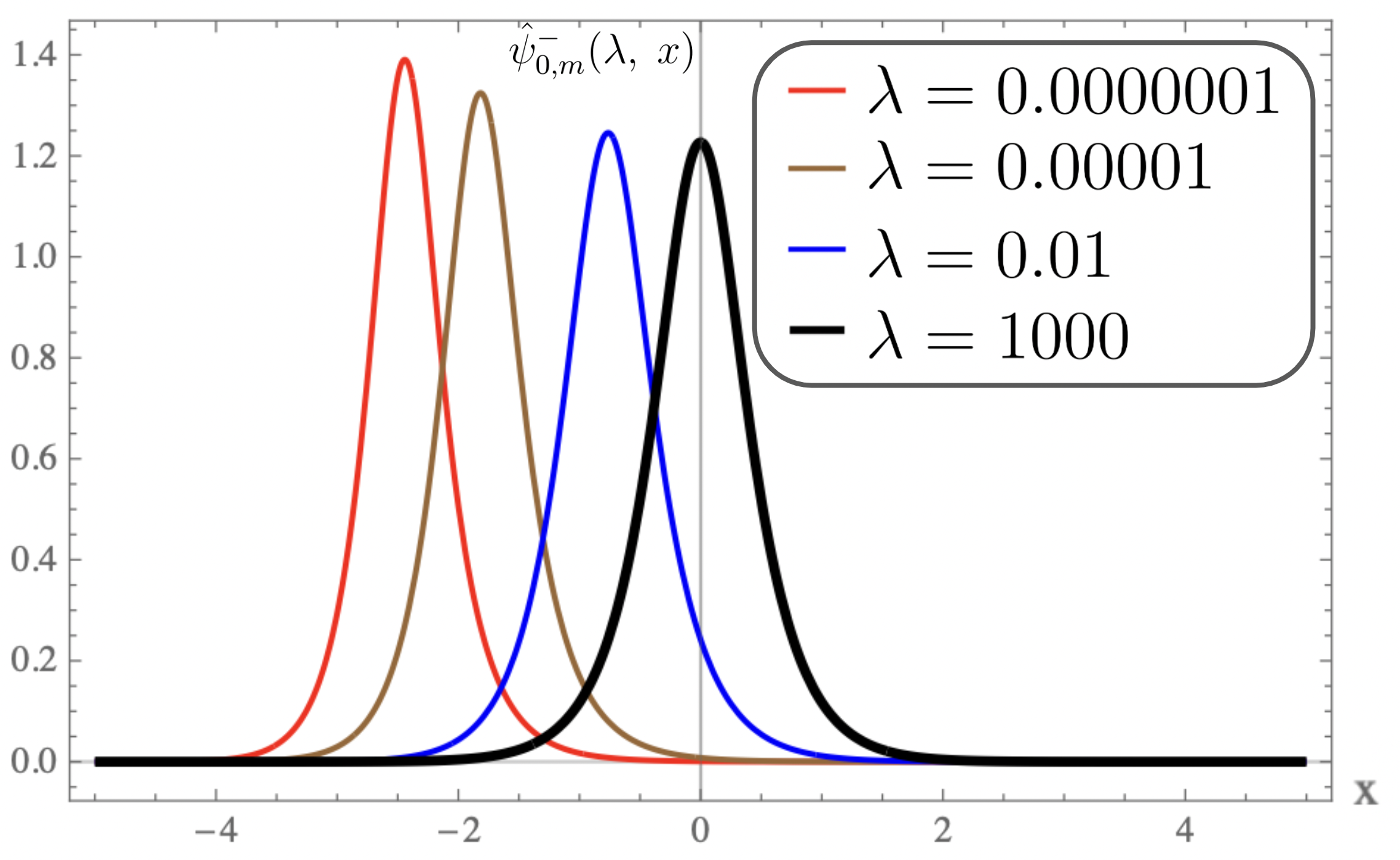}}
\caption{Plot of isospectral potential \(\hat{V}^-_4(\lambda, x)\)  and its ground state eigenfunction \( \hat{\psi}^-_{0,4}(\lambda,x)\) as a function of \( x \) for various \( \lambda \) when \(m=4\).}\label{figure-iso-4}
\end{figure}

\section{One parameter family of REHO potentials}
The One parameter ($\lambda$) family of strictly isospectral potentials 
corresponding to \( V^-_m(x) \) are easily obtained using (\ref{iso-gs}-\ref{iso-es}) as
\begin{equation}
\hat{V}^-_m(\lambda,x)=V^-_m(x)-2 \frac{d^2}{dx^2}\ln\left[\mathcal{I}_m(x)+\lambda\right],\label{ISO-V}
\end{equation}
where the integral \( \mathcal{I}_m(x) \) is calculated using (\ref{RE-GS}) as
\begin{equation}
 \mathcal{I}_m(x)= \left( \frac{2^m m!}{\sqrt{\pi}}\right)\int^x_{-\infty} \left[ \frac{e^{- \frac{{x'}^2}{2}}}{\mathcal{H}_m(x')}\right]^2 dx'\label{I}.
\end{equation}\\
The expressions for \( \mathcal{I}_m(x) \) and \( \hat{V}^-_m(\lambda,x) \) for
\( m \) equal to \( 0,\;2\text{ and }4 \) are given in table-\ref{tab-yi} and 
table-\ref{tab-iso-v} respectively in terms of the error function 
\( \text{erf}(x) \) defined as
\begin{equation*}
\text{erf}(x)=1-\text{erfc}(x),\qquad \text{erfc}(x)= \frac{2}{\sqrt{\pi}}
\int^{\infty}_x e^{-t^2}dt\,.
\end{equation*}\\
The normalized ground state wave function \( \hat{\psi}^-_{0,m}(\lambda, x) \) 
is obtained using (\ref{iso-gs}), (\ref{RE-GS}) and (\ref{I}) as
\begin{equation}
\hat{\psi}^-_{0,m}(\lambda, x)= \sqrt{\lambda(1+\lambda)}\;
\frac{\psi^-_{0,m}(x)}{\mathcal{I}_m(x)+\lambda}\,.\label{ISO-GS}
\end{equation}

\begin{table}[htp]
\centering
\setlength{\tabcolsep}{2em}
\begin{tabularx}{\columnwidth}{@{}>{\bfseries}l X X l @{}}
\toprule
m & \(y^m_{n+1}(x),\qquad n=0,1,2\cdots \)&\( \mathcal{I}_m(x) \)\\
\toprule
0& \( H_{n+1}(x) \) &\(\frac{1}{2} (\text{erf}(x)+1)  \)\\
2& \( 2 \left(2 x^2+1\right) H_{n+1}(x)+8 x H_n(x) \) &\( \frac{1}{2} \left(\text{erf}(x)+\frac{2 e^{-x^2} x}{\sqrt{\pi } \left(2 x^2+1\right)}+1\right) \)\\
4& \( 4 \left( 4x^4+12x^2+3\right) H_{n+1}(x)+ \left(64 x^3+96x\right) H_n(x)\) &\(\frac{1}{2} \left(\text{erf}(x)+\frac{2 e^{-x^2} x \left(2 x^2+5\right)}{\sqrt{\pi } \left(4 \left(x^2+3\right) x^2+3\right)}+1\right)  \)\\
\bottomrule
\end{tabularx}
\caption{ Exceptional Hermite polynomials \( y^m_{n+1} \) and  \(\mathcal{I}_m(x)\) for different \( m \).}
\label{tab-yi}
\end{table}


\begin{table}[htp]
\centering
\setlength{\tabcolsep}{1em}
\begin{tabularx}{\columnwidth}{@{}>{\bfseries}l X l @{}}
\toprule
m &  \( \hat{V}^-_m(\lambda, x) \)\\
\toprule
0& \(x^2-2+ \frac{e^{-2 x^2} \left(8 \sqrt{\pi } e^{x^2} x (\text{erf}(x)+2 \lambda +1)+8\right)}{\pi  (\text{erf}(x)+2 \lambda +1)^2}\)\\
2& \(x^2-2 +\frac{8 \left(\pi  e^{2 x^2} \left(2 x^2-1\right) (\text{erf}(x)+2 \lambda +1)^2+6 \sqrt{\pi } e^{x^2} x (\text{erf}(x)+2 \lambda +1)+4\right)}{\pi  e^{2 x^2} \left(2 x^2+1\right)^2 (\text{erf}(x)+2 \lambda +1)^2+4 \sqrt{\pi } e^{x^2} \left(2 x^3+x\right) (\text{erf}(x)+2 \lambda +1)+4 x^2}\)\\
4& \( x^2-2+\frac{16 \left(\pi  e^{2 x^2} \left(8 x^6+12 x^4+18 x^2-9\right) (\text{erf}(x)+2 \lambda +1)^2+16 \sqrt{\pi } e^{x^2} x \left(x^4+x^2+3\right) (\text{erf}(x)+2 \lambda +1)+4 \left(2 x^4+x^2+8\right)\right)}{\left(\sqrt{\pi } e^{x^2} \left(4 \left(x^2+3\right) x^2+3\right) (\text{erf}(x)+2 \lambda +1)+2 x \left(2 x^2+5\right)\right)^2}\)\\
\bottomrule
\end{tabularx}
\caption{Expression of isospectral potential \(\hat{V}^-_m(x)\) for different \( m \).}
\label{tab-iso-v}
\end{table}
The normalized excited states eigenfunction of \( \hat{V}^-_m(\lambda,x) \), using (\ref{iso-es}) are given by
\begin{align}
\hat{\psi}^-_{n+1,m}(\lambda, x)&=\frac{1}{E^-_{n+1,m}} \hat{A}^{\dagger}A\psi^-_{n+1,m}(x)\label{ISO-ES},\qquad n=0,1,2,\cdots,\\
&=\frac{e^{-\frac{x^2}{2}}}{\sqrt{2(n+m+1)2^nn!\sqrt{\pi}}} \left( \frac{y^m_{n+1}(x)}{\mathcal{H}_m(x)}+    H_n(x)\frac{d}{dx}\ln\left[\mathcal{I}_m(x)+\lambda\right]  \right)\nonumber
\end{align}
and 
\begin{equation*}
\hat{A}^{\dagger}=-\frac{d}{dx}-\frac{d}{dx}\ln\left[\frac{\psi^-_{0,m}(x)}{\mathcal{I}_m(x)+\lambda}\right],\qquad A=\frac{d}{dx}-\frac{d}{dx}\ln [\psi^-_{0,m}(x)].
\end{equation*}
The expression of \(  \hat{\psi}^-_{n+1,m}(x)\) for \( m \) equal to 
\( 0,\;2\text{ and }4 \) are tabulated in table-\ref{tab-iso-es} . As mentioned
above, the spectra of \( \hat{H}^-_m \), defined using (\ref{Hiso-}), is strictly isospectral to \( H^-_m \) spectra and is given by Eq. (\ref{1}).

\begin{table}[htp]
\centering
\setlength{\tabcolsep}{1em}
\begin{tabularx}{\columnwidth}{@{}>{\bfseries}l  l @{}}
\toprule
m &  \( \hat{\psi}^-_{n+1,m}(\lambda, x) ,\qquad n=0,1,2,\cdots\)\\
\toprule
0 & \(\frac{e^{-\frac{x^2}{2}}}{\sqrt{2(n+1)2^nn!\sqrt{\pi}}} \left( \frac{y^0_{n+1}(x)}{1}+    \frac{2   H_n(x)}{e^{x^2}\sqrt{\pi } (\text{erf}(x)+2 \lambda +1)}  \right)\)\\
2 & \(\frac{e^{-\frac{x^2}{2}}}{\sqrt{2(n+3)2^nn!\sqrt{\pi}}} \left( \frac{y^2_{n+1}(x)}{4x^2+2}+   \frac{4   H_n(x)}{\left(2 x^2+1\right) \left(\sqrt{\pi } e^{x^2} \left(2 x^2+1\right) (\text{erf}(x)+2 \lambda +1)+2 x\right)}   \right) \)\\
4 & \(\frac{e^{-\frac{x^2}{2}}}{\sqrt{2(n+5)2^nn!\sqrt{\pi}}} \left( \frac{y^4_{n+1}(x)}{16x^4+48x^2+12}+   \frac{48  H_n(x)}{\sqrt{\pi } e^{x^2} \left(4 \left(x^2+3\right) x^2+3\right)^2 (\text{erf}(x)+2 \lambda +1)+2 x \left(8 x^6+44 x^4+66 x^2+15\right)}    \right) \)\\
\bottomrule
\end{tabularx}
\caption{Expression of excited state wavefunction \( \hat{\psi}^-_{n+1,m}(x) \) for different \( m \).}
\label{tab-iso-es}
\end{table}

For \( m \) equal to \( 0,\;2\text{ and } 4 \), the plots of the strictly 
isospectral potentials for positive and negative \( \lambda \) as a function of \( x \)
are shown in Fig-(\ref{figure-iso-0}.a, \ref{figure-iso-2}.a, 
\ref{figure-iso-4}.a) and Fig-(\ref{figure-iso-0}.b, \ref{figure-iso-2}.b, 
\ref{figure-iso-4}.b) respectively. Pursey and Abraham-Moses potential plots as a function of \( x \) are shown in Fig-(\ref{figure-iso-0}.c, \ref{figure-iso-2}.c, 
\ref{figure-iso-4}.c). The ground-state wavefunction plots as a function of \( x \) for various \( \lambda \) are shown in Fig-(\ref{figure-iso-0}.d, \ref{figure-iso-2}.d, 
\ref{figure-iso-4}.d).
From the figures one observes that the eigen functions and the strictly 
isospectral potentials become sharper with increasing \(m  \). 
In the limit \( \lambda \) approaching to \( \pm \infty \) the potential 
\( \hat{V}^-_m(\lambda,\;x) \) approaches to \( V^-_m(x) \). 
Also notice that the potential starts developing a minimum when \( \lambda \) 
decreases from 
\( \infty \) to zero and the attractive potential well shifts towards 
\( -\infty \) and finally vanishes when \(  \lambda\) equals zero. There is a 
loss of bound state and the corresponding potential is called the Pursey 
potential \( V^{P}_m(x) \). An analogous situation occurs in the limit 
\( \lambda=-1 \) and the potential is called the Abraham-Moses potential 
\( V^{AM}_m(x) \).

\subsection{The Pursey and The Abraham-Moses Potentials}
The Pursey and The Abraham-Moses Potentials are obtained from (\ref{ISO-ES}) 
by substituting \( \lambda=0 \) and \( \lambda=-1 \) respectively. In this case
as mentioned above, one looses a bound state and the spectrum is identical to
that of $H^{+}$, i.e.
\begin{equation*}
\hat{E}^{P}_{n,m}=\hat{E}^{AM}_{n,m}=E^+_{n,m}\,,
\end{equation*}
where
\begin{equation*}
E^+_{n,m}=2(n+m+1),\qquad n=0,1,2,3,\cdots
\end{equation*}
The normalized eigenfunctions of \( \hat{V}^{P}_m(x) \) are given by
\begin{align}
\hat{\psi}^{P}_{n,m}(x)&=\frac{1}{E^-_{n+1,m}} \left(-\frac{d}{dx}-\frac{d}{dx}\ln\left[\frac{\psi^-_{0,m}(x)}{\mathcal{I}_m(x)}\right]\right)A\psi^-_{n+1,m}(x),\qquad n=0,1,2,\cdots,\label{P}.
\end{align}
Similarly, the normalized eigenfunctions of \( \hat{V}^{AM}_m(x) \) is given by
\begin{align}
\hat{\psi}^{AM}_{n,m}(x)&=\frac{1}{E^-_{n+1,m}} \left(-\frac{d}{dx}-\frac{d}{dx}\ln\left[\frac{\psi^-_{0,m}(x)}{\mathcal{I}_m(x)-1}\right]\right)A\psi^-_{n+1,m}(x),\qquad n=0,1,2,\cdots,\label{AM}.
\end{align}

\begin{figure}
\hfill
\subfigure[\bf Plot of first excited state eigenfunction \( \hat{\psi}^-_{1,2}(\lambda,\;x) \) corresponding to potential \( \hat{V}^-_2(\lambda,x) \) as a function of \( x \) when \( \lambda \) approaches 0.]{\includegraphics[height=5.8cm, width=7.8cm]{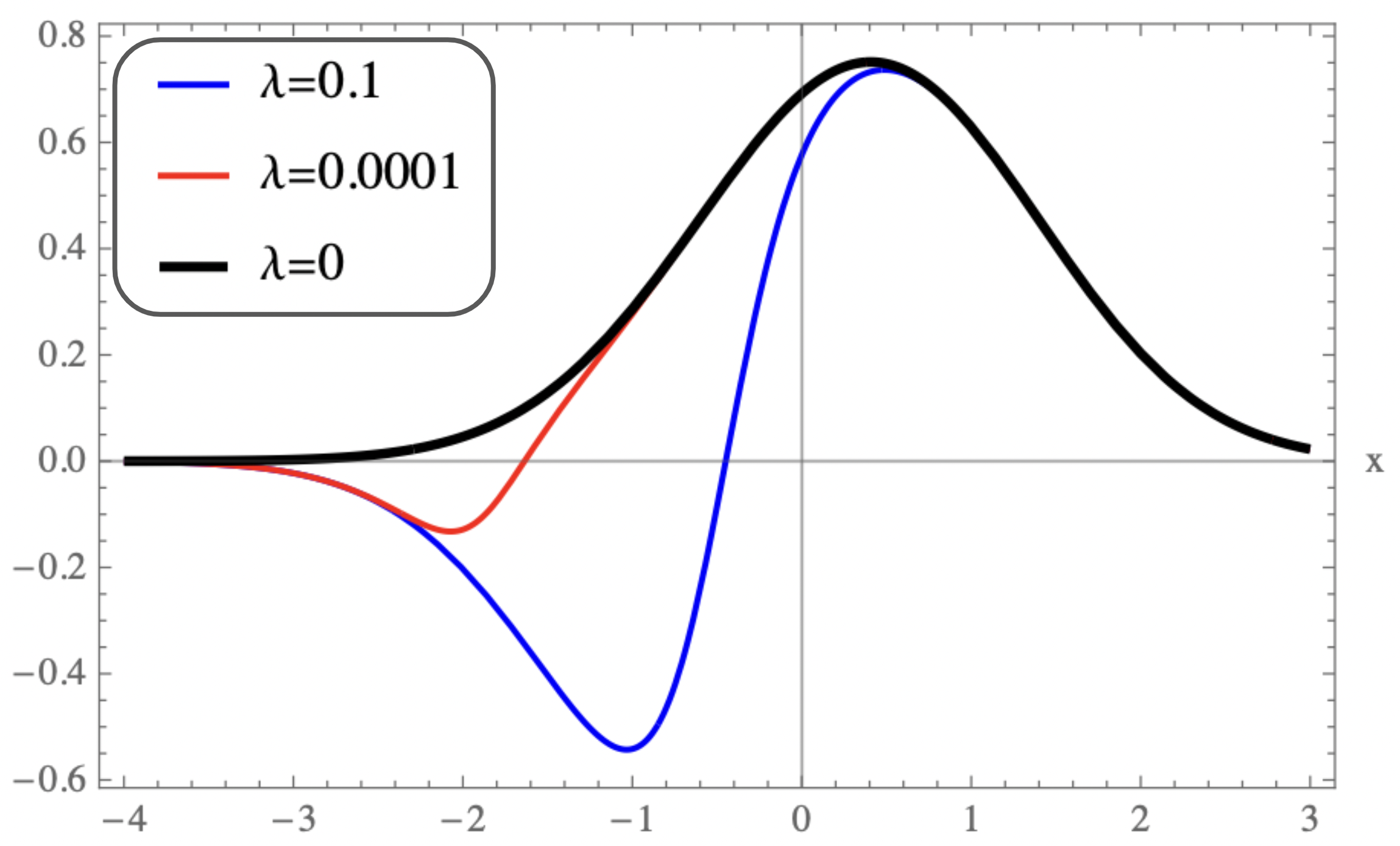}}
\hfill
\subfigure[\bf Plot of ground state wavefunction of Pursey and AM potentials as a function of \( x \) when \(m=0\) (Black curve), \(m=2\)(Red curve) and \(m=4\)(Blue curve).]{\includegraphics[height=5.8cm, width=7.8cm]{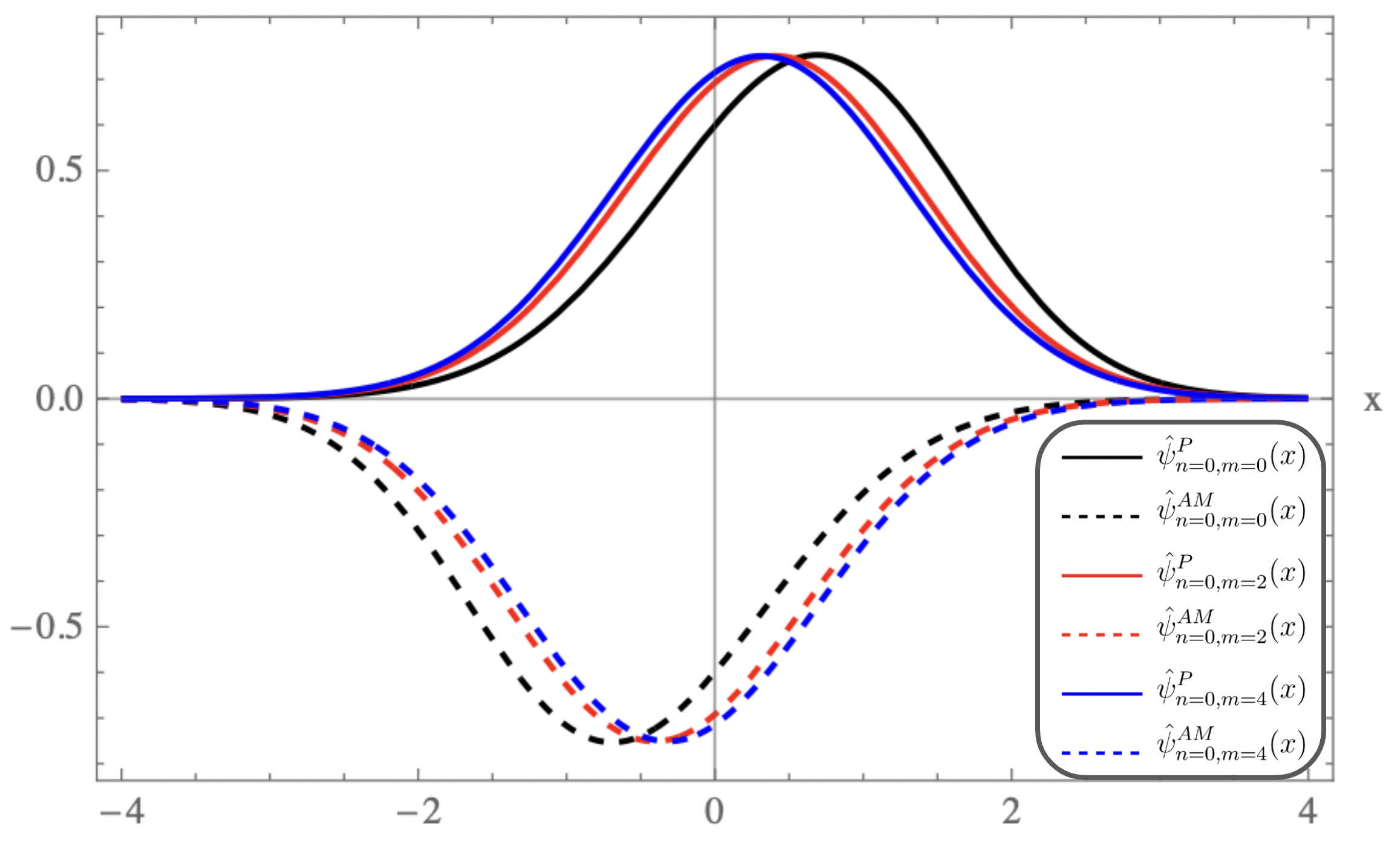}}
\hfill
\caption{Dashing curve represents Pursey eigenfunction and dashed curve represents the AM eigenfunction.}
\label{figure-pursey-am}
\end{figure}

The expression of \(  \hat{\psi}^{P}_{n,m}(x)\) and \( \hat{\psi}^{AM}_{n,m}(x) \) for \( m \) equal to \( 0,\;2\text{ and }4 \) are obtained from table-\ref{tab-iso-es} by substituting \( \lambda \) equal to \( 0 \) and \( -1 \) respectively.\\
The ground state wavefunction plot for Pursey and AM potentials for various \( m \) are shown in figure-\ref{figure-pursey-am}.b and transition in wavefunction shape as \( \lambda \) approaches zero is shown in figure-\ref{figure-pursey-am}.a.

\section{Calculation of Uncertainty $\Delta x \Delta p$}
The Heisenberg Uncertainty relation for position, \( x \), and momentum, \( p \), is defined as 
\begin{equation}
\Delta x\Delta p=\sqrt{\left(<x^2>\;-\;<x>^2\right)\;\left(<p^2>\;-\;<p>^2\right)},
\end{equation}
where, the angle bracket, \( <> \), represents expectation value in a given wavefunction basis.

\subsection{Uncertainty relation for REHO potential}
For the REHO potentials the expectation value of the position, \( x \), and 
the momentum, \( p \), in the ground as well as the excited states is zero as 
the integrand is an odd function in the calculation of \( <x> \) while the 
expectation value of \( <p> \) is zero since the eigenfunctions are real. The 
uncertainty relation therefore takes the simpler form
\( \Delta x\Delta p =\sqrt{<x^2><p^2>}\)

The expectation values and hence the uncertainty values can be calculated by
using Eqs. (\ref{RE-GS}). It is observed that \( \Delta x =\Delta p\) for \( m=0 \) but the same is not true for \( m\neq 0 \).  The uncertainty values for $m = 0, 1,2$ and $n = -1, 0, 1$ (i.e. ground, first and second excited states) are shown in table-\ref{tab-un-re}.

\begin{figure}
\subfigure[\bf Uncertainty at ground state for versus \( m \).]{\includegraphics[height=5cm, width=8cm]{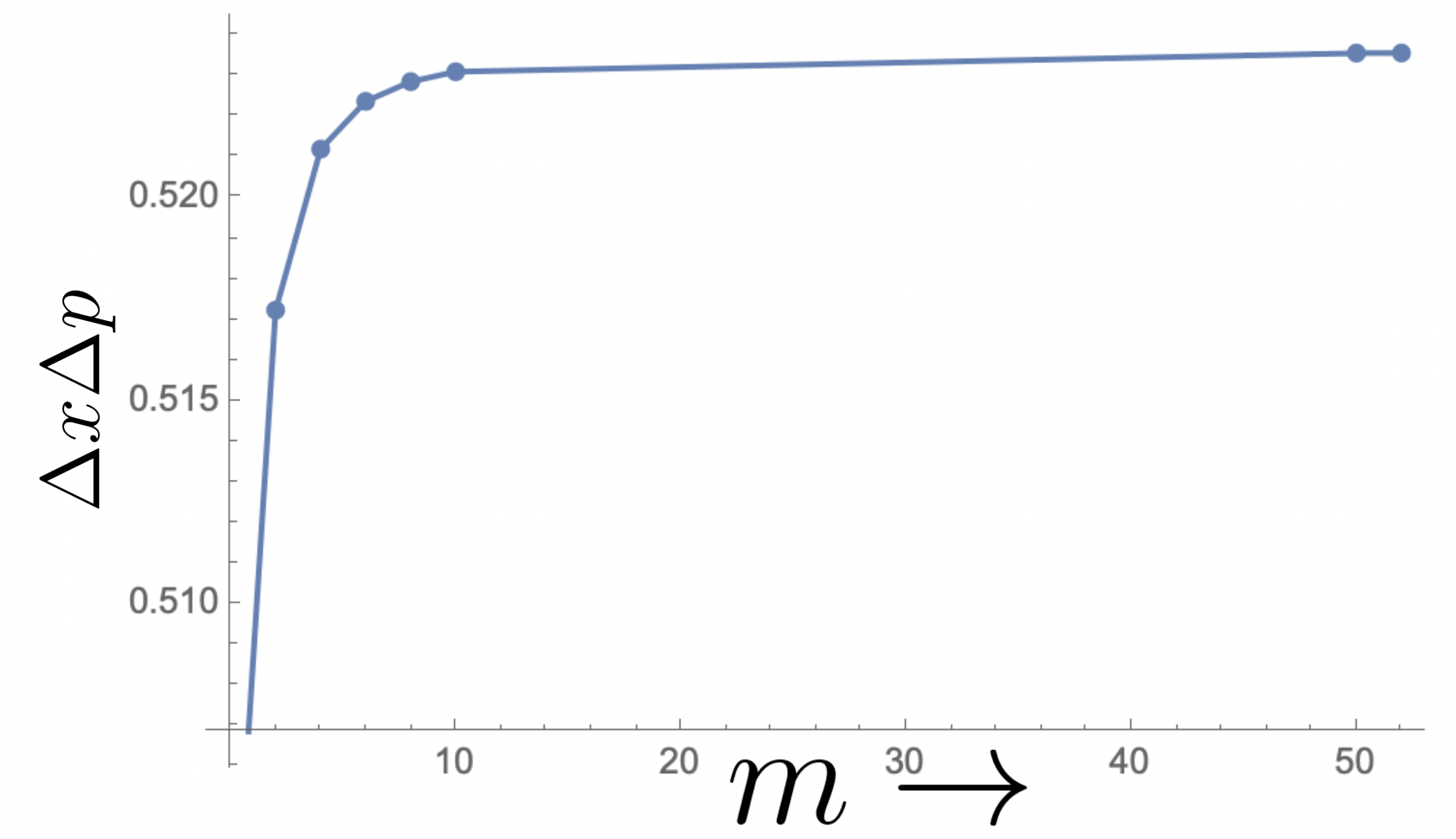}}
\hfill
\subfigure[\bf Uncertainty at different excited state for \( m \) equal to 0, 8 and 52.]{\includegraphics[height=5cm, width=8cm]{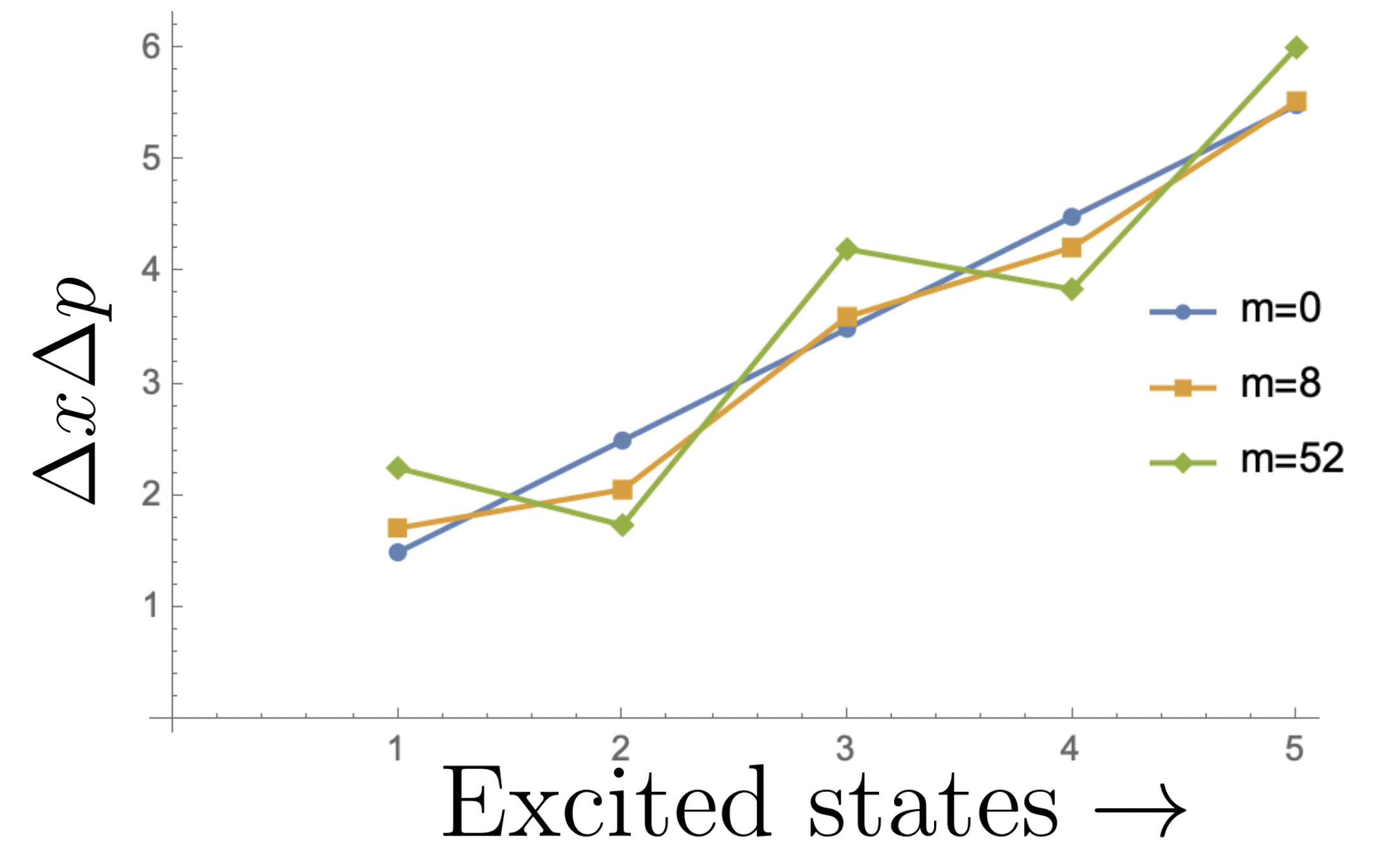}}
\caption{Uncertainty trends with increasing \( m \) and \( n \).}
\label{fig-un-re}
\end{figure}

\begin{table}[htp]
\centering
\setlength{\tabcolsep}{1em}
\begin{tabularx}{\columnwidth}{@{}>{\bfseries}l X X l @{}}

\toprule
m & \( n=-1 \) & \( n=0 \)&\( n=1 \)\\
\toprule
0&\( 0.5\)&\(1.5 \)&\(2.5 \)\\
2&\( \approx 0.5172\)&\(\approx1.554 \)&\( \approx2.3412\)\\
4&\( \approx 0.5212\)&\( \approx1.6152\)&\( \approx2.2102\)\\
\bottomrule

\end{tabularx}
\caption{ Uncertainty relation \( \Delta x \Delta p \) for potential \( V^-_m(x) \) in ground state, first excited state and second excited state for \( m \) equal to \( 0,\;2,\text{ and }4 \).}
\label{tab-un-re}
\end{table}
It can be seen from figure-\ref{fig-un-re}.a that for the ground state, the 
uncertainty value increases as \( m \) increases from $0$ and then flattens 
out around $m = 10$. On the other hand, in the different excited states the
Uncertainty values show peculair behaviour. In particular, while it decreases 
with increasing \( m \) for any given even excited states but it increases with
increasing $m$ for any given odd excited states. This peculiar behaviour in 
the uncertainty may be attributed to the even and odd degree of exceptional 
Hermite polynomials for a given \( n \). The figure-\ref{fig-un-re}.b shows the
variation of uncertainty with different excite states for \( m \) equal to 0, 8 and 52.
\newpage

\subsection{Uncertainty relation for one parameter family of REHO potential}
The expectation value of momentum, \( <p> \), is again zero as the 
eigenfunctions are real. The uncertainty in \( \Delta p \) is, therefore, 
\(\sqrt{<p^2>} \). 
The expectation values of \( x \), \( x^2 \) for various values of \( m \) and 
\( \lambda \) are given in Appendix A. 
The uncertainty relation at ground-state for various values of \( m \) and \( \lambda \) is given in table-\ref{tab-un-iso}.

\begin{table}[htp]
\centering
\setlength{\tabcolsep}{1em}
\begin{tabularx}{\columnwidth}{@{}>{\bfseries}l X X X X X l @{}}

\toprule
m/ \( \lambda=\)& \(1\times 10^{-12} \) & \(1\times10^{-8} \)& \(1\times10^{-5} \) & \(1\times10^{-3}\)& \(1\times10^{-1} \)& \( 1\times10^{2} \) \\
\toprule
0 & \(  \approx 0.5202\) & \( \approx 0.5281 \)& \( \approx 0.5367 \) & \(  \approx 0.5590\)& \( \approx 0.5455 \)& \( \approx 0.5000 \) \\
2 & \( \approx 0.5223 \) & \( \approx 0.5270 \)& \( \approx 0.5285 \) & \( \approx 0.5266 \)& \( \approx 0.5193 \)& \( \approx 0.5172 \) \\
4 & \( \approx 0.5228 \) & \( \approx 0.5252 \)& \( \approx 0.5246 \) & \( \approx 0.5232 \)& \( \approx0.5215 \) & \( \approx 0.5212 \)\\
\bottomrule
\end{tabularx}
\caption{ Uncertainty relation \( \Delta x \Delta p \) for potential \( \hat{V}^-_m(\lambda,x) \) in ground state for \( m \) equal to \( 0,\;2,\text{ and }4 \).}
\label{tab-un-iso}
\end{table}

\begin{figure}
\subfigure[\bf m=0 case.]{\includegraphics[height=5cm, width=8cm]{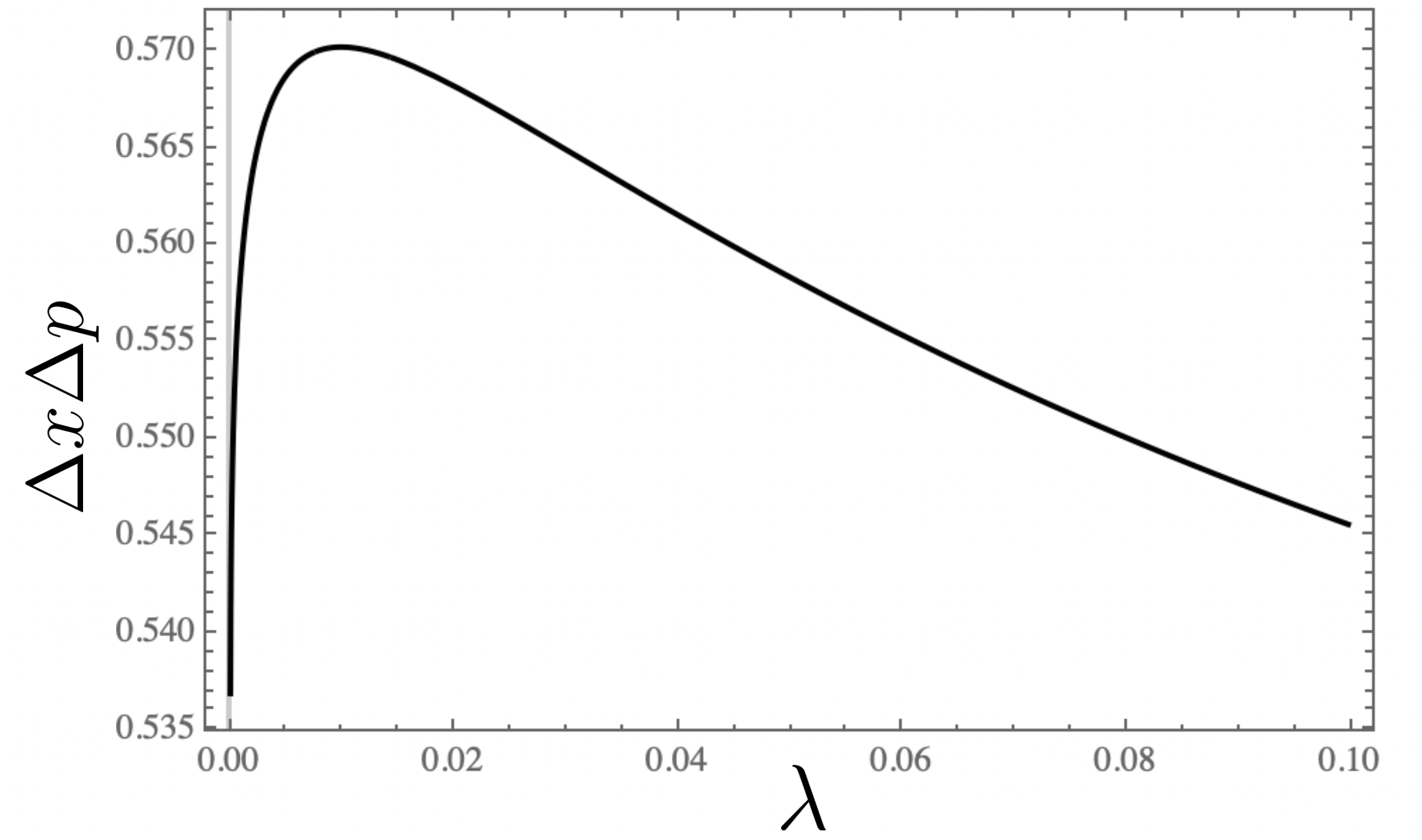}}
\hfill
\subfigure[\bf m=2 case.]{\includegraphics[height=5cm, width=8cm]{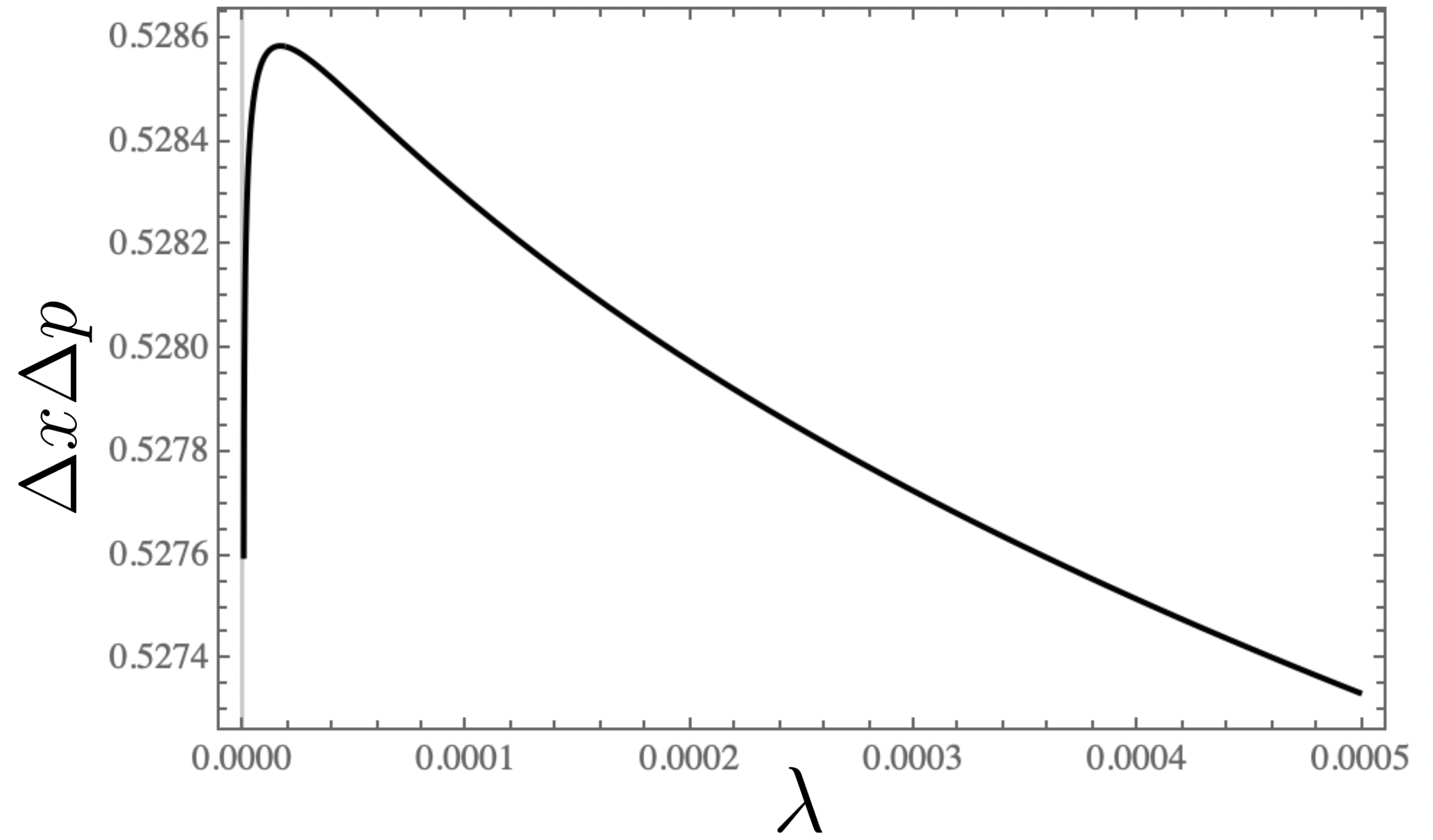}}
\caption{Uncertainty versus positive \(\lambda\) graph for \( m=0 \) and \( m=2 \).}\label{figure-uncertainty}
\end{figure}

When the values of \( \lambda (>0)\) in the table-\ref{tab-un-iso} are replaced by \( -|\lambda+1| \) the uncertainty relation \( \Delta x \Delta p \) remains unchanged and it corresponds to the Abraham Moses potential.
The plot of uncertainty versus \( \lambda \) for \( m=0 \) and \( m=2 \) is shown in figure-\ref{figure-uncertainty}.a and \ref{figure-uncertainty}.b respectively.
The peak of uncertainty curve rapidly moves towards origin with increasing \( m \). 

\subsubsection{Uncertainty relations for Pursey and Abraham Moses Potentials}
The expectation value of \( x \) and \( x^2 \) are calculated in Appendix C for
various values of \( m \). It is interesting to note that the expectation value of 
\( x \) is equal and opposite in the case of the Pursey and the AM Potentials
while the expectation value of $x^2$ and $p^2$ is the same in both the cases.

\begin{table}[htp]
\centering
\setlength{\tabcolsep}{1em}
\begin{tabularx}{\columnwidth}{@{}>{\bfseries}l X X X X l @{}}
\toprule
m/n & 0 & 1& 2&3&10\\
\toprule
0&\(\approx 0.50184 \)&\( \approx 1.4879\)&\(\approx 2.4894 \)&\( \approx 3.4905\)&\(\approx 10.4947 \)\\
2&\( \approx 0.50015\)&\(\approx 1.4990 \)&\( \approx 2.4984\)&\( \approx 3.4980\)&\(\approx 10.4977 \)\\
4&\( \approx 0.50004\)&\(\approx 1.4997 \)&\(\approx 2.4994 \)&\( \approx 3.4992\)&\( \approx 10.4987\)\\
\bottomrule
\end{tabularx}
\caption{ Uncertainty relation \( \Delta x \Delta p \) for potential \( \hat{V}^{P}_m(x) \) or \( \hat{V}^{AM}_m(x) \) for various \( n \) and for \( m \) equal to \( 0,\;2,\text{ and }4 \).}
\label{tab-un-pam}
\end{table}

As a result the uncertainty value corresponding to a given \( m \) is the 
same for the Pursey and the AM potentials. The uncertainty value for the Pursey and the AM potentials at the ground and the different excited-states is given in table-\ref{tab-un-pam}.  The uncertainty decreases at ground state with increasing \( m \) and becomes asymptotic to \( 0.5 \). On contrast it is seen that uncertainty is lesser than QHO (\( m=0 \)) uncertainty for excited states.

\section{Summary and discussion}
In this manuscript, we consider the rationally extended one dimensional 
harmonic oscillator potential associated with exceptional $X_m$-Hermite 
polynomials. Unlike the one dimensional oscillator, for nonzero $m$, while the
energy gap between the excited states is $2$ units, the energy gap between 
the groundstate and first excited state is \( m+1 \). Using the idea of SQM, we 
obtained one parameter ($\lambda$) family of strictly isospectral potentials 
corresponding to the REHO potential as well as the corresponding 
eigenfunctions. As a special case, as $\lambda$ approaches $0$ or $-1$, we 
obtained the rationally extended Pursey and the AM potentials respectively as 
well as their eigenfunctions. 
Further, we calculated the Heisenberg uncertainty relations $\Delta x \Delta p$
for the REHO and as well as the corresponding strictly isopectral one parameter
($\lambda$) family of potentials. In addition, we also calculated the 
uncertainity relation for the corresponding Pursey and AM potentials. In the 
case of the REHO, we showed that the ground state 
uncertainity increases as \(m\) increases. In the case of the strictly 
isospectral one parameter family of potentials, one finds that the uncertainty
relation depends on $m$ as well as $\lambda$. Remarkably, we find that for any
$m$, the uncertainty relation is the same for the Pursey and the AM potentials. 

There are several open problems. For example, In this paper we have only 
studied the
potentials with even codimension $m$ and obtained the corresponding strictly 
isospectral one parameter family. Can one similarly obtain the corresponding
one parameter family of potentials in case the codimension $m$ is odd. Further,
apart from the one dimensional
oscillator, are there other rationally extended potentials for which the
eigenfunctions can be expressed in terms of exceptional Hermite polynomials? 
Another obvious question is whether there are exceptional Legendre polynomials
and if yes can one discover new  potentials whose eigenfunctions can be 
expressed in terms of exceptional Legendre polynomials? We hope to study some
of these problems in coming days.

{\bf Acknowledgements}\\
AK is gratefel to Indian National Science Academy (INSA) for awarding INSA 
Honarary Scientist position at Savitribai Phule Pune University. One of us (RK) is grateful to Ian Marquette for useful comments.\\

{\bf Appendix A. Expectation value of \( x \) and \( x^2 \) for one parameter 
family of REHO} \\
The expectation value of \( x \) and \( x^2 \) for various values of positive \( \lambda \) are calculated using (\ref{ISO-GS}) as:
\begin{equation*}
<x> =
\begin{cases}
\approx -3.0028\|_{m=0}\;\approx -2.2259\|_{m=2}\;\approx -1.8057\|_{m=4}&\text{, when $\lambda=0.00001$}\\
\approx -2.1553\|_{m=0}\;\approx -1.4304\|_{m=2}\;\approx -1.1187\|_{m=4}&\text{, when $\lambda=0.001$}\\
\approx -0.9133\|_{m=0}\;\approx -0.5202\|_{m=2}\;\approx -0.3955\|_{m=4}&\text{, when $\lambda=0.1$}\\
\approx -0.0039\|_{m=0}\;\approx -0.0022\|_{m=2}\;\approx -0.0016\|_{m=4}&\text{, when $\lambda=100$}\\
\approx -0.0004\|_{m=0}\;\approx -0.0002\|_{m=2}\;\approx -0.0002\|_{m=4}&\text{, when $\lambda=1000$}\\\end{cases}
\end{equation*}

When the values of \( \lambda (>0) \) in the above table are replaced by 
\( -|\lambda+1| \) the magnitude of the values of \( x \) remain unchanged but 
the sign reverses.

\begin{equation*}
<x^2> =
\begin{cases}
\approx 9.1024\|_{m=0}\;\approx 5.0384\|_{m=2}\;\approx 3.3276\|_{m=4}&\text{, when $\lambda=0.00001$}\\
\approx 4.8071\|_{m=0}\;\approx 2.1621\|_{m=2}\;\approx 1.3309\|_{m=4}&\text{, when $\lambda=0.001$}\\
\approx 1.2338\|_{m=0}\;\approx 0.4198\|_{m=2}\;\approx 0.2446\|_{m=4}&\text{, when $\lambda=0.1$}\\
\approx 0.5000\|_{m=0}\;\approx 0.1556\|_{m=2}\;\approx 0.0896\|_{m=4}&\text{, when $\lambda=100$}\\
\approx 0.5\|_{m=0}\;\approx 0.1556\|_{m=2}\;\approx 0.0896\|_{m=4}&\text{, when $\lambda=1000$}\\\end{cases}
\end{equation*}
When the values of \( \lambda (>0) \) in the above table are replaced by \( -|\lambda+1| \) the expectation values of \( x^2 \) remains unchanged.
Similarly, the expectation value of  \( p^2 \) can be calculated.\\

{\bf Appendix B. Expectation value of \( x \) and \( x^2 \) for Puresey and 
Abraham Moses Potentials} \\
The expectation value and uncertainty relation can be calculated from (\ref{P}) and (\ref{AM}).\\

The expectation value of \( x \) and \( x^2 \) for various values of \( m \) are as follows:
\begin{equation*}
<x> =
\begin{cases}
\approx 0.6386\|_{\text{Pursey}}\;\approx -0.6386\|_{\text{AM}}&\text{, when $m=0$}\\
\approx 0.3928\|_{\text{Pursey}}\;\approx -0.3928\|_{\text{AM}}&\text{, when $m=2$}\\
\approx 0.3087\|_{\text{Pursey}}\;\approx -0.3087\|_{\text{AM}}&\text{, when $m=4$}\\
\end{cases}
\end{equation*}

\begin{equation*}
<x^2> =
\begin{cases}
\approx 0.9043\|_{\text{Pursey}}\;\approx 0.9043\|_{\text{AM}}&\text{, when $m=0$}\\
\approx 0.6539\|_{\text{Pursey}}\;\approx 0.6539\|_{\text{AM}}&\text{, when $m=2$}\\
\approx 0.5952\|_{\text{Pursey}}\;\approx 0.5952\|_{\text{AM}}&\text{, when $m=4$}\\
\end{cases}
\end{equation*}

\end{document}